\documentclass[iop,tighten]{emulateapj}
\usepackage{ifpdf} 
\pdfoutput=1  % arxiv.org needs this in the first 5 lines
\usepackage{graphicx} 
\usepackage{cmap} 
\usepackage{mathptmx} 
\usepackage[T1]{fontenc}
\usepackage{natbib}
\usepackage[usenames, dvipsnames]{xcolor} 
\definecolor{lightgreen}{HTML}{228B22} 
\definecolor{darkgreen}{HTML}{006400} 
\usepackage[english]{babel}
\usepackage[colorlinks, bookmarks, breaklinks, pdftitle={Towards Precision Supermassive Black Hole Masses using Megamaser Disks},pdfauthor={Remco C. E. van den Bosch}]{hyperref} \hypersetup{linkcolor=black,citecolor=black,filecolor=black,urlcolor=darkgreen,pdfborderstyle={/S/U}} 
\usepackage{nicefrac} 
\usepackage{relsize} 
\usepackage[activate=true,kerning=true,tracking=true,spacing=true,babel,final]{microtype} 
\usepackage{hyphenat}
\usepackage{type1cm} 
\usepackage{lettrine}

\def\gtrsim{\mathrel{\hbox{\rlap{\hbox{\lower4pt\hbox{$\sim$}}}\hbox {\raise2pt\hbox{$>$}}}}} 
\newcommand{\kms}{km~s\ensuremath{^{-1}}} 
\newcommand{\msun}{\ensuremath{M_{\odot}}} 
\newcommand{\mbh}{\ensuremath{M_\mathrm{BH}}} 
\newcommand{\sigmastar}{\ensuremath{\sigma_{\ast}}} 
\newcommand{\msigma}{\ensuremath{M_{\mathrm{BH}}-\sigmastar}} 
\newcommand{\MsigmaC}{\ensuremath{M_{\mathrm{BH}}-\sigma_\mathbf{c}}} 
\newcommand{\oiii}{[\ion{O}{3}]} 

\slugcomment{Second minor revision submitted to {\it The Astrophysical Journal} on 23/12/2015. Original version: 7/8/2015} \shorttitle{{\it No Maser Disks in Elliptical Galaxies}} \shortauthors{van den Bosch, Greene, et al.}

\begin{document} 
\title{Towards Precision Supermassive Black Hole Masses using Megamaser Disks}

\author{\href{http://mpia.de/~bosch}{Remco C. E. van den Bosch}\altaffilmark{1},
Jenny E. Greene\altaffilmark{2}, \\
James A. Braatz\altaffilmark{3}, 
Anca Constantin\altaffilmark{4}, 
Cheng-Yu Kuo\altaffilmark{5}}

\altaffiltext{1}{\href{http://mpia.de/~bosch}{Max-Planck Institut f\"ur Astronomie}, K\"onigstuhl 17, D-69117 Heidelberg, Germany, \href{mailto:bosch@mpia.de}{bosch@mpia.de}} \altaffiltext{2}{Department of Astrophysics, Princeton University, Princeton, NJ 08544, USA, \href{mailto:jgreene@astro.princeton.edu}{jgreene@astro.princeton.edu}} \altaffiltext{3}{National Radio Astronomy Observatory, 520 Edgemont Road, Charlottesville, VA 22903, USA} \altaffiltext{4}{Department of Physics and Astronomy, James Madison University, Harrisonburg, VA 22807, USA} \altaffiltext{5}{Physics Department, National Sun Yat-Sen University, No. 70, Lien-Hai Rd, Kaosiung City 80424, Taiwan, R.O.C}

\begin{abstract} Megamaser disks provide the most precise and accurate extragalactic supermassive black hole masses. Here we describe a search for megamasers in nearby galaxies using the Green Bank Telescope (GBT). We focus on galaxies where we believe that we can resolve the gravitational sphere of influence of the black hole and derive a stellar or gas dynamical measurement with optical or NIR observations. Since there are only a handful of super massive black holes (SMBH) that have direct black hole mass measurements from more than one method, even a single galaxy with a megamaser disk and a stellar dynamical black hole mass would provide necessary checks on the stellar dynamical methods. We targeted 87 objects from the Hobby-Eberly Telescope Massive Galaxy Survey, and detected no new maser disks. Most of the targeted objects are elliptical galaxies with typical stellar velocity dispersions of 250~\kms\ and distances within 130~Mpc. We discuss the implications of our non-detections, whether they imply a threshold X-ray luminosity required for masing, or possibly reflect the difficulty of maintaining a masing disk around much more massive ($\gtrsim 10^8$~\msun) black holes at low Eddington ratio. Given the power of maser disks at probing black hole accretion and demographics, we suggest that future maser searches should endeavour to remove remaining sample biases, in order to sort out the importance of these covariant effects. 
\end{abstract}

%\keywords{super massive black holes: mass measurements, scaling relations, host galaxies}

\section{Introduction} \label{sec:Introduction}

\lettrine[slope=-2pt,nindent=-2pt,lines=2]{W}{ater} megamasers at 22 GHz are detected in $\sim 3\%$ of local Seyfert 2 and LINER galaxies that have been searched. In roughly one-third of these, characteristic red- and blue-shifted components indicate that they arise from sub-pc scales in a geometrically thin accretion disk around a weakly accreting super massive black hole \citep[BH;][]{lo2005, pesce15}. The best-known example is NGC4258 \citep{miyoshietal1995}, but there are now more than 15 megamaser disk galaxies known that show clean Keplerian rotation, and 34 total megamaser disk galaxies \citep{zhuetal2011,pesce15}.

Despite their small numbers, water megamasers have disproportionate scientific impact. Thousands of galaxies have been searched for maser activity, largely by the \href{https://safe.nrao.edu/wiki/bin/view/Main/MegamaserProjectSurvey}{Megamaser Cosmology Project} \citep[MCP, ][]{reidetal2009,braatzetal2010}. By measuring the acceleration of the systemic water maser features, it is possible to measure direct geometric distances to these galaxies \citep[e.g.,][]{humphreysetal2013}. The main goal of the MCP is to garner a precise and independent measurement of $H_0$ \citep{reidetal2013, kuoetal2013, kuoetal2015}.

Modeling of the near-Keplerian rotation of the maser spots also yields very precise and accurate BH masses, with uncertainties of $\sim 10\%$ that are dominated by the uncertainty in the galaxy distance. The deviations from Keplerian rotation are so small that it is possible to rule out astrophysical alternatives to supermassive BHs in these systems \citep[e.g.,][]{kuoetal2011}. These black hole masses have much smaller uncertainties than those obtained by other methods and are thus ideally suited for cross validation of other black hole mass measurement techniques.

To date, there are only a handful of cross-checks on dynamical black hole mass measurements. The different methods often do not yield consistent masses \citep[e.g.][]{walshetal2013, walshetal12a, onkenetal14}. Even stellar dynamical methods do not always agree; see, for instance, NGC3379 \citep{shapiroetal06a, vandenboschdezeeuw2010}, NGC1399 \citep[][]{Houghtonetal06, gebhardtetal07} and NGC4258 \citep{siopisetal2009, drehmer15}. There is no single culprit for these discrepancies, because different methods and practices are used as gas and stars probe the potential in different ways. The inhomogeneity of the mass measurements, and the low number statistics, make it impossible to quantify the systematic uncertainty. 

So far, the only megamaser disk galaxy with either a stellar or gas dynamical BH mass measurement is NGC4258. For this object, the state-of-the-art orbit-based models based on \textit{HST} optical long-slit spectroscopy find a black hole mass that is 15\% lower than the megamaser-derived mass \citep{siopisetal2009}, whereas simpler Jeans models with adaptive optics near-infrared integral field spectroscopy find a value that is 25\% too high \citep{drehmer15}. Furthermore, the sizes of the quoted uncertainties are such that neither of these measurements are within 3$\sigma$ of the maser value. This is typical for the other cross-checks also. It thus appears that systematic unknowns dominate the uncertainty. More comparisons are clearly needed to quantify and understand these discrepancies.

Further cross-checks would also provide information on the intrinsic scatter of BH-galaxy scaling relations. The widely used \msigma\ relation correlates the velocity dispersion of the host galaxy with the black hole mass. This relation (and many others) have an intrinsic scatter of $\sim 0.4$~dex (or more) \citep[e.g.,][]{beifiori12}. The intrinsic scatter as a function of galaxy properties should contain important clues on the origin of the scaling relations \citep[e.g.,][]{robertsonetal2006,peng2007,jahnkemaccio2011}.  However, we cannot measure the intrinsic scatter robustly until we understand the underlying systematic uncertainties in the dynamical BH masses. Perhaps the size of the measured scatter is artificially inflated by the systematic uncertainties of the black hole mass measurements? 

With current instrumentation it will be very difficult to attempt any cross-calibration using the megamaser black hole masses. Most of the maser galaxies are so distant that it is impossible to spatially resolve the stellar motions in the region of the galaxy where the gravity of the BH dominates (known as the gravitational sphere of influence, see \S~\ref{sec:Samples}). Thanks to the superior spatial resolution of the VLBI observations, the masers can probe well within the sphere of influence for BHs down to $10^6$~\msun\ even at 100 Mpc. Using stellar or gas dynamical techniques it is not yet possible to measure a $10^6$~\msun\ at distances beyond $2.5$ Mpc. 

The known masers with spheres of influence large enough ($>0.1$\arcsec) for a robust measurement from a dynamical technique are Circinus, NGC4945 and NGC4258. The latter was discussed above. The other two masers show non-disk components, which makes their interpretation more ambiguous than the `clean` masers \citep{pesce15}.  Measuring their black hole mass with a dynamical technique would be a good test of the maser mass. However these objects are not ideal for understanding the systematic uncertainty of the dynamical BH masses.

Thus, the only way to increase the sample of galaxies with direct comparisons between stellar/gas dynamical and megamaser disk masses is to find a new megamaser disk in a galaxy where we can resolve the gravitational sphere of influence.  That goal motivated the megamaser search presented here. We focus on galaxies with the largest spheres of influence on the sky, to boost the chances that we can get a megamaser and stellar/gas dynamical mass in the same object. 

Previous surveys have looked for megamaser disks in a variety of galaxy types. See \cite{henkeletal2005} for a nice summary. The majority of searches to date have focused on known obscured active galaxies, to maximize the chance to catch an edge-on disk. These active galaxies have been selected in a variety of ways. The majority of galaxies were observed with the sensitive Robert C. Byrd Green Bank Telescope (GBT), which has a gain of $\sim$1.9 K/Jy at K-band. The MCP has compiled a public list (\href{https://safe.nrao.edu/wiki/bin/view/Main/MegamaserProjectSurvey}{weblink}) of $\sim$3400 galaxies that have been searched for megamaser with the GBT. Recent work \citep{zhuetal2011,zhangetal2010,zhang12,constantin2012} investigates the detection fraction of megamaser galaxies as a function of various galaxy parameters. Because of our goal to maximize the likely BH sphere of influence, our survey is dominated by relatively massive elliptical galaxies relative to previous searches, providing new constraints on the physical properties that lead some galaxies to mase. 

Here we present our unsuccessful attempt to identify new masing disks in nearby galaxies. In \S \ref{sec:Samples} we discuss the sample selection and observations of our maser search. We discuss overall maser detection fractions in \S \ref{sec:Results}, and discuss the implications of our results for the physics of and detection of megamaser disks in \S \ref{sec:Discussion}, and conclude in \S \ref{sec:summary}. We adopt a Hubble constant $H_0$ of 70 \kms. This value is consistent with all the published values based on geometric distance determinations of megamaser disk galaxies \citep[e.g.][]{reidetal2013,kuoetal2013}.

\section{Our Search for Megamaser Disk Galaxies} \label{sec:Samples} 

Our goal is to find new megamasers in nearby galaxies for cross calibration of BH mass measurements. Thus we are searching for megamaser disks in galaxies that are near enough to measure the dynamical impact of the BH on the surrounding stars or gas. We use the Hobby-Eberly Telescope Massive Galaxy Survey \citep[HETMGS, ][]{vandenboschetal2015} as our parent sample. The HETMGS uses several different target selections to find all galaxies that are suitable for stellar- and gas-dynamical supermassive BH mass measurements in the optical or near-infrared. This is achieved by maximizing the apparent size of the sphere-of-influence of the SMBHs in the target galaxies. The HETMGS survey contains 1022 galaxies and observed nearly all possible targets for dynamical black hole masses measurements in the North ($-11<\delta<73^\circ$). It is thus an ideal basis for a nearby maser search.

\subsection{Search Criteria}

We apply two selection criteria. First, we use the stellar velocity dispersion measurement from HETMGS to roughly estimate whether or not a dynamical BH measurement will be possible. Specifically, the gravitational sphere of influence of the BH ($\theta_{\rm I} \propto \frac{GM_{\rm BH}}{D \sigma_{\ast}^2}$; where $D$ is the distance) needs to be spatially resolved ($\theta_{\rm I} > 0.06\arcsec$) in order to probe regions near the black hole. Since we have no measurement of the BH mass, we use \mbh$\propto \sigma^4_{\ast}$ from \citet{gultekinetal2009} as a proxy\footnote{The HETMGS only measured the velocity dispersion $\sigma_c$ inside a central $\sim$2\arcsec\ aperture. Hence an \MsigmaC\ is presented in \citet{vandenboschetal2015} which would be more appropriate to predict black hole masses. The traditional \msigma\ uses the dispersion measured within the galaxy effective radius $\sigma_\mathbf{e}$, which is typically much larger than 2 arcseconds. Like many more recent relations \citep{mcconnellma2013,kormendyho2013}, the relation based on $\sigma_c$ is steeper than the \mbh$\propto \sigma^4_{\ast}$ from \citet{gultekinetal2009} that is adopted in this work. The relation with $\sigma_c$ would yield 631 objects with $\theta_{\rm I} > 0.06\arcsec$, but this fit was not available when we performed our search.}. There are 369 such galaxies in the HETMGS. \label{sec:soidefinition}

\begin{figure}
\begin{center}
\includegraphics[width=0.45
\textwidth]{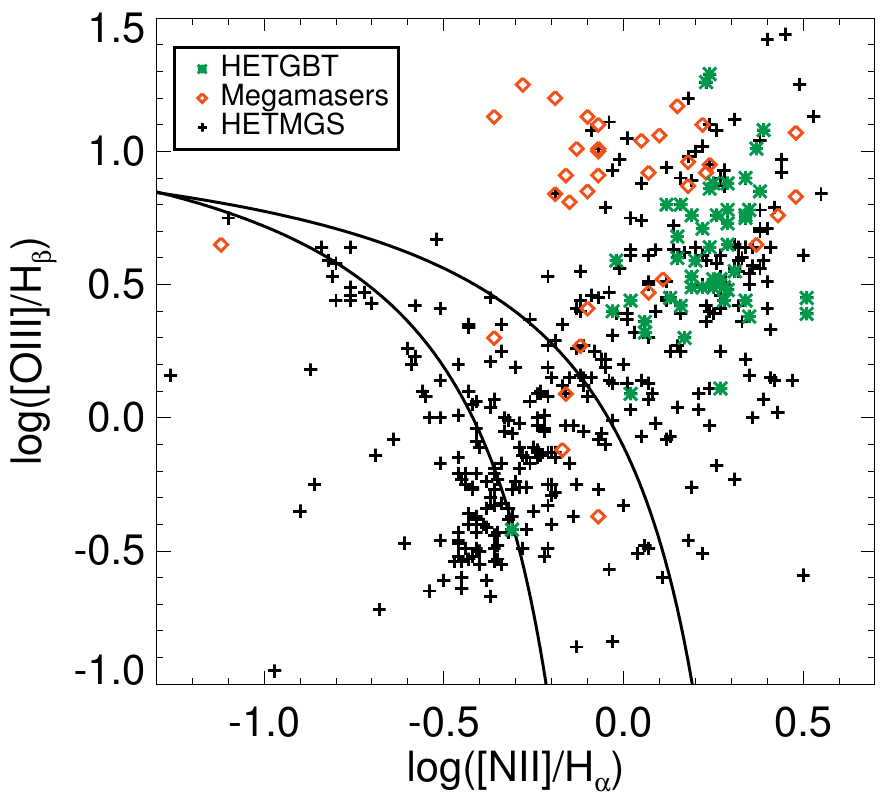}
\end{center}
\vskip -3mm
\figcaption[]{ The BPT diagnostic diagram used to identify active galaxies for this survey. Shown are the emission lines ratios [N{\tiny II}]/H$_\alpha$ $~\lambda 6583$ and [O{\tiny III}]/H$_\beta$$~\lambda 5007$ from the HETMGS survey. Green crosses represent the HETGBT sample that we observed. Red diamonds are known megamasers in the HETMGS. For reference, the remaining HETMGS sources in which emission lines are detected are shown as black crosses. The upper and lower black lines delineate the empirical and theoretical divide \citep{kewley06} between star-formation and other emission (AGN, shocks). Our targets are selected to have large spheres of influence and non-starforming nuclear emission, i.e to fall on the upper-right side of the divide. \label{fig:bpt}}
\end{figure}

Second, we only select galaxies with optical signs of nuclear activity. Ideally, we would search all 369 galaxies with $\theta_{\rm I} > 0.06\arcsec$, but in order to boost the probability of finding a maser disk, we restricted our attention to galaxies with optical signatures of accretion. We use the HETMGS spectra\footnote{Because the HET's optical path changes during each of the observations, an absolute flux calibration was not performed on the HET data. The spectra were corrected in a relative sense for the spectral response. However no absolute line luminosities are available for the HETMGS spectroscopy.} to measure strong emission lines ratios [N{\tiny II}]$~\lambda 6583$/H$_\alpha$ and \mbox{[O {\tiny III}]$~\lambda 5007$/H$_\beta$}. Standard \citet*[BPT,][]{baldwinetal1981} diagnostic diagrams are used to isolate the active galactic nuclei (AGN; Fig.\ \ref{fig:bpt}) from the galaxies with (nuclear) star-formation. Selecting only the AGN further reduces the target list to 121 objects. For this search, we did not distinguish between Seyferts and low-ionization nuclear emission region LINERs \citep{heckmanetal1981}, but see section~\ref{sec:liners}.

Finally, we checked for overlap between the HETMGS and the GBT searches. After removing all galaxies previously searched for megamaser emission, 93 galaxies remained in our target list. This last cut removed almost all of the bright AGN from our target list. See Fig.~\ref{fig:bpt} for the distribution in the BPT diagram of the observed galaxies.

%%%%%%%%%%%%%%%%%%%%%%%%%%%%%%%%%%%%%%%%%%%%%%%%%%%%%%%%%%%%%%%%%%%%%%
%%%%%%%%%%%%%%%%%%%%%%%%%%%%%%%%%%%%%%%%%%%%%%%%%%%%%%%%%%%%%%%%%%%%%
\begin{figure*}
\begin{center}
\vbox{ 
\hskip 5mm
\includegraphics[width=0.9\textwidth]{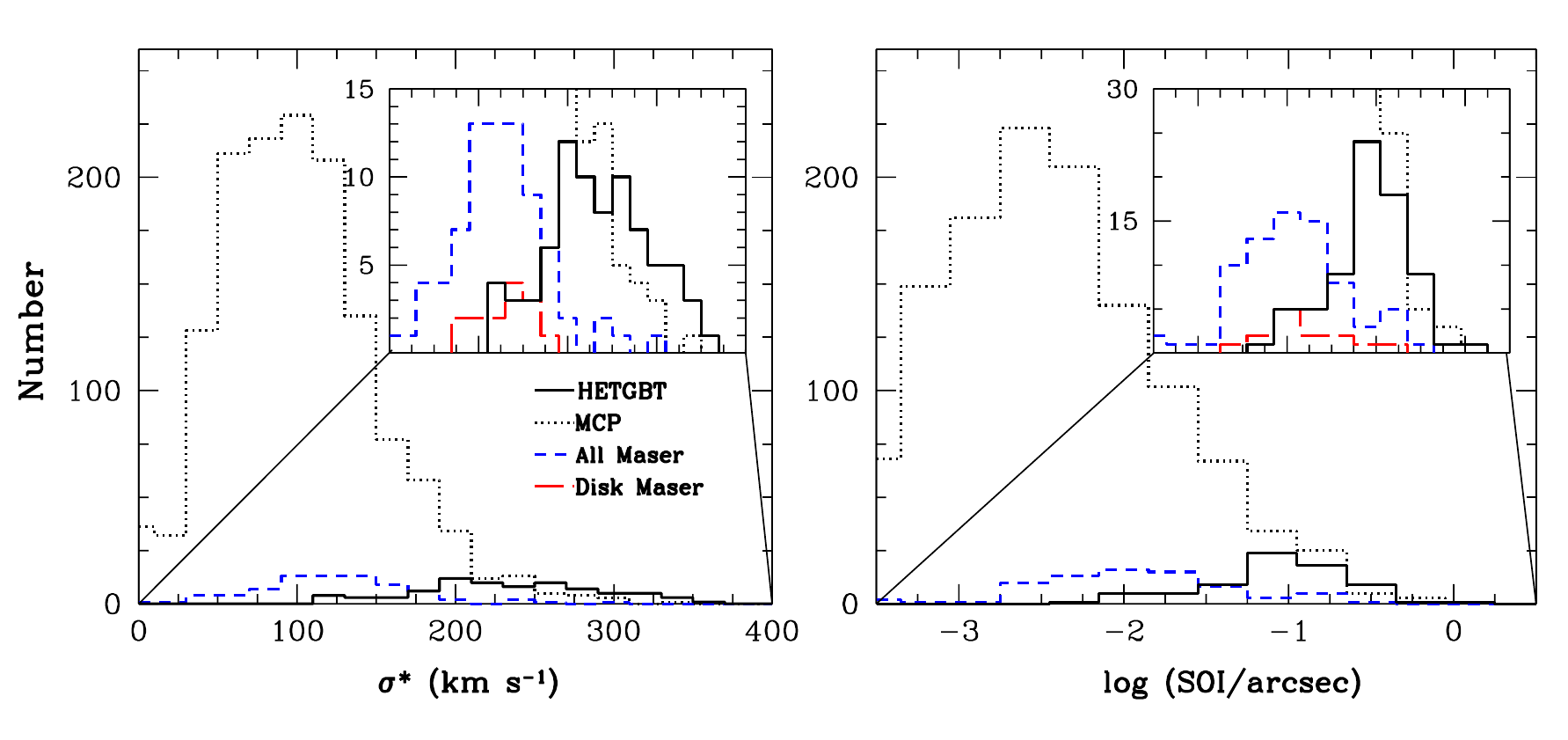} }

\figcaption[]{ {\it Left}: Histogram of the stellar velocity dispersions. We show our HETGBT sample that we searched for masers in black solid, the MCP non-detections in dotted, all MCP masing galaxies in blue dashed and the maser disk galaxies in red long-dashed. The zoomed in region has the same horizontal span but shows that we more than double the number of $\sigma_{\ast} > 250$ \kms\ galaxies. {\it Right}: Histogram of the spheres-of-influence of known megamaser disks (red), the HETGBT (black) and the known non-detections from the MCP (dotted). Galaxies with large spheres-of-influence are inherently rare. \label{fig:sig_hist}} 

\end{center}
\end{figure*}
%%%%%%%%%%%%%%%%%%%%%%%%%%%%%%%%%%%%%%%%%%%%%%%%%%%%%%%%%%%%%%%%%%%%%%

\subsection{GBT Observations and Data Reduction} \label{sec:Observations}

We conducted our survey using the K-Band Focal Plane Array (KFPA) and the GBT Spectrometer. We used 2 beams of the KFPA in the same mode used by the MCP. We used 2 IF's of 200 MHz bandwidth offset by 180 MHz, for a total coverage of 380 MHz. Each window covers 2800 \kms\ and, with overlap, the total coverage is 5100 \kms. We do not know a priori the velocity extent of the putative megamaser disks in these more massive galaxies. However, if we had detected only systemic masers in any system, we would have expanded the search to higher velocities. We used a total-power observing mode, nodding the telescope to alternate the target galaxy between the two beams on a 2.5-minute interval. The receiver pointed off source is used as the reference beam to measure the background. The reference beam spectrum is smoothed with a kernel of 16 channels, to increase its signal-to-noise. In good weather (T$_{sys} \sim 40$~K) and integrating 10 minutes per galaxy, we achieved $\sim$ 4 mJy rms per channel after Hanning smoothing. The pointing corrections, done roughly each hour, were typically 5\arcsec\ or better, and the flux calibration is accurate to about 20\%. The final velocity resolution and channel spacing is 0.4 and 0.3 km s$^{-1}$. This observing setup is  sufficient to identify any megamasers that can be imaged in follow-up observations with a single VLBI track.

We observed 87 galaxies using 21.5 hours during a visit to the NRAO's GBT\footnote{The National Radio Astronomy Observatory is a facility of the National Science Foundation operated under cooperative agreement by Associated Universities, Inc.} from 6 through 15 November 2012 as program \href{https://library.nrao.edu/proposals/catalog/6588}{GBT/12B-052}. See Table~1 with observations of this HETGBT sample. Apart from galaxies from our target list, the 87 galaxies include five filler objects and the control maser NGC6240. The filler objects have properties close to our selection criteria and are included in our HETGBT sample. At our maximum distance of 130 Mpc, the 3$\sigma$ luminosity limit is $\sim 0.7$ L$_{\odot}$, while the least luminous known H$_2$O masing disk is in NGC2273, which has an isotropic luminosity of $23$ L$_{\odot}$, so we are not limited by sensitivity.

\subsection{Sample Properties}

The sample has a median distance of 74 Mpc and a median stellar velocity dispersion of $250$~km~s$^{-1}$. The majority of the sample galaxies are elliptical galaxies (see next paragraph), with likely BH masses in excess of $10^8$~\msun. Since the sphere of influence depends steeply on $\sigma_{\ast}$, our sample is heavily skewed towards high-dispersion galaxies. As shown in Figure \ref{fig:sig_hist}, our program more than doubles the number of galaxies with $\sigma > 250$ \kms\ that have been surveyed for maser emission with the GBT. The observed galaxies have larger inferred spheres-of-influence than the galaxies previously searched as well. As shown in Figure \ref{fig:sig_hist}, this survey doubles the number of galaxies searched with $\theta_{\rm I} > 0.06\arcsec$.

Only 44 objects have a morphological type $T$ in Hyperleda \citep{patureletal2003}. Almost all of these are ellipticals: $T=-3.2\pm1.9$. The only known spiral in our sample is IC0356. In general, we do not have deep and reliable imaging with which to determine Hubble types. Thus, to put the morphological distribution of our sample in context, we plot the mass-size relation of all galaxies searched for masers by the GBT. Ellipticals obey a tight scaling between size and mass and are the smallest (densest) galaxies at a given stellar mass. As Figure~\ref{fig:sizes} shows, the HETMGS galaxies that we searched preferentially probe dense elliptical galaxies, while previous searches were overwhelmingly dominated by spiral galaxies. Our search probes the locus of early-type galaxies and adds to a part of parameter space that has not yet been exhaustively searched.

%%%%%%%%%%%%%%%%%%%%%%%%%%%%%%%%%%%%%%%%%%%%%%%%%%%%%%%%%%%%%%%%%%%%%
\vbox{ \vskip +1mm \hskip -2mm 
\includegraphics[width=0.45\textwidth]{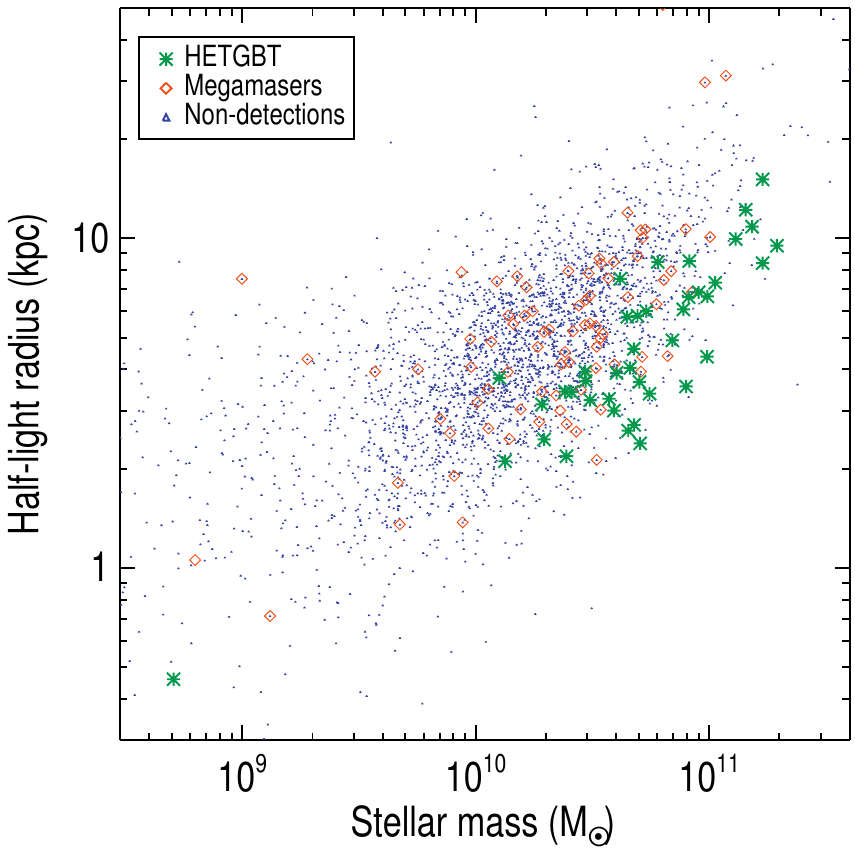} }
\vskip -1mm 

\figcaption[]{Size-mass diagram of searched galaxies, using half-light radii and stellar masses from the NASA Sloan Atlas by \protect{\citet{blanton09}} for all objects where SDSS photometry is available. There are 2927 GBT non-detections (blue dots), 83 masers (red diamonds) and 41 HETGBT galaxies that we searched (green crosses) with SDSS photometry. The MCP non-detections cover the main population of late-type galaxies, whereas our search probed the locus of massive early-type galaxies, that were hitherto not sampled.\label{fig:sizes}}

\vskip 6mm
%%%%%%%%%%%%%%%%%%%%%%%%%%%%%%%%%%%%%%%%%%%%%%%%%%%%%%%%%%%%%%%%%%%%%

There was a large survey of elliptical galaxies carried out by \citet{henkel98}, but it was focused on luminous radio galaxies, so selected in a different way from this search. Indeed, there are only a few known megamasers in elliptical galaxies: NGC1052 \citep{braatzetal1994,tarchietal2003}, 3C403 \citep{tarchietal2007}, Centaurus~A \citep{ott13}, possibly NGC2960 \citep{kormendyho2013}. Object TXS2226-184 is most likely not an early-type galaxy \citep{falcke00}. About half of our sample is detected in NVSS (NRAO VLA Sky Survey), with a median luminosity of 10 mJy \citep{condon98}.

\begin{figure*}
\begin{center}
\vbox{ \vskip +1mm %\hskip +10mm 
\includegraphics[height=6.6cm]{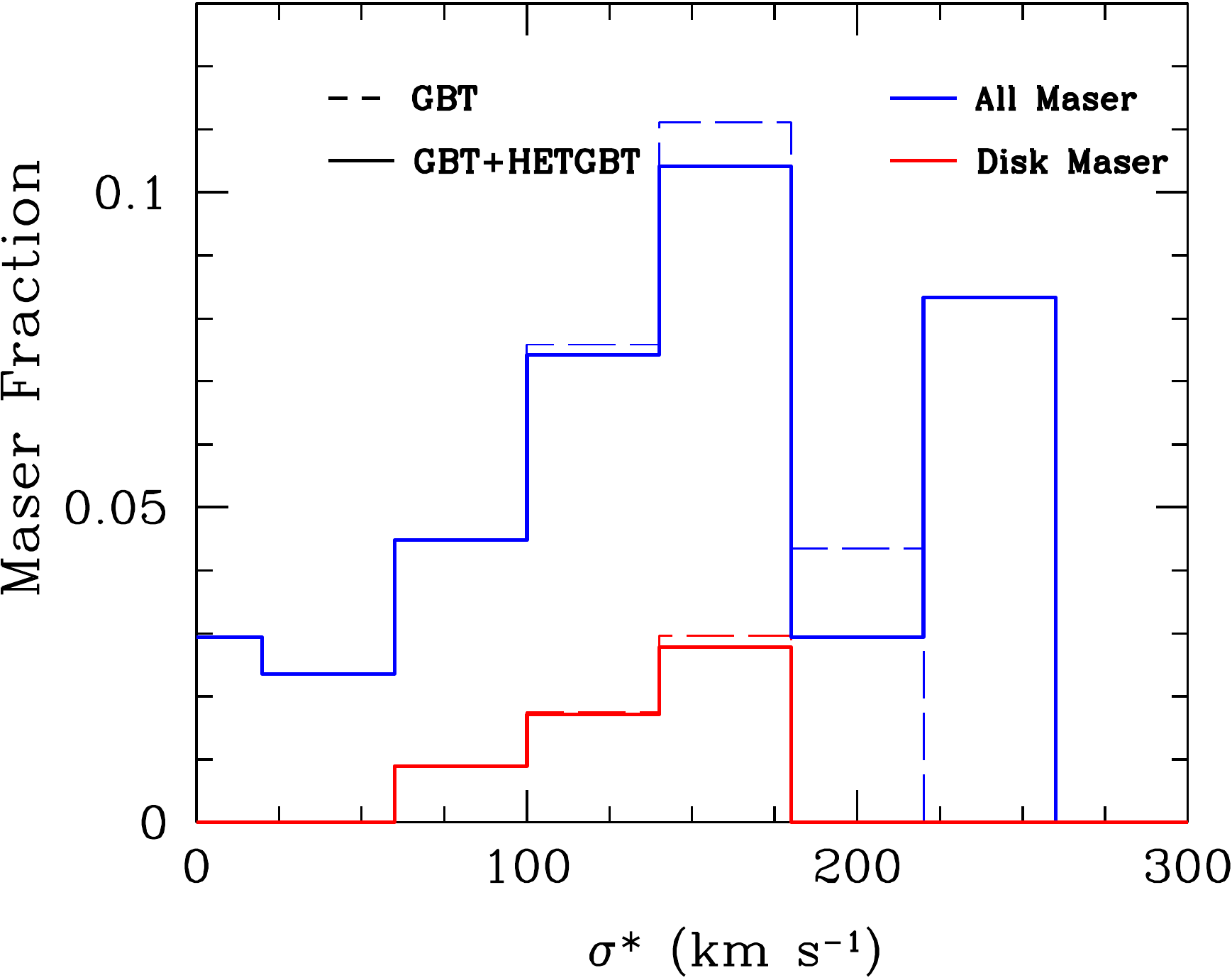}   \hskip 5mm 
\includegraphics[height=6.6cm]{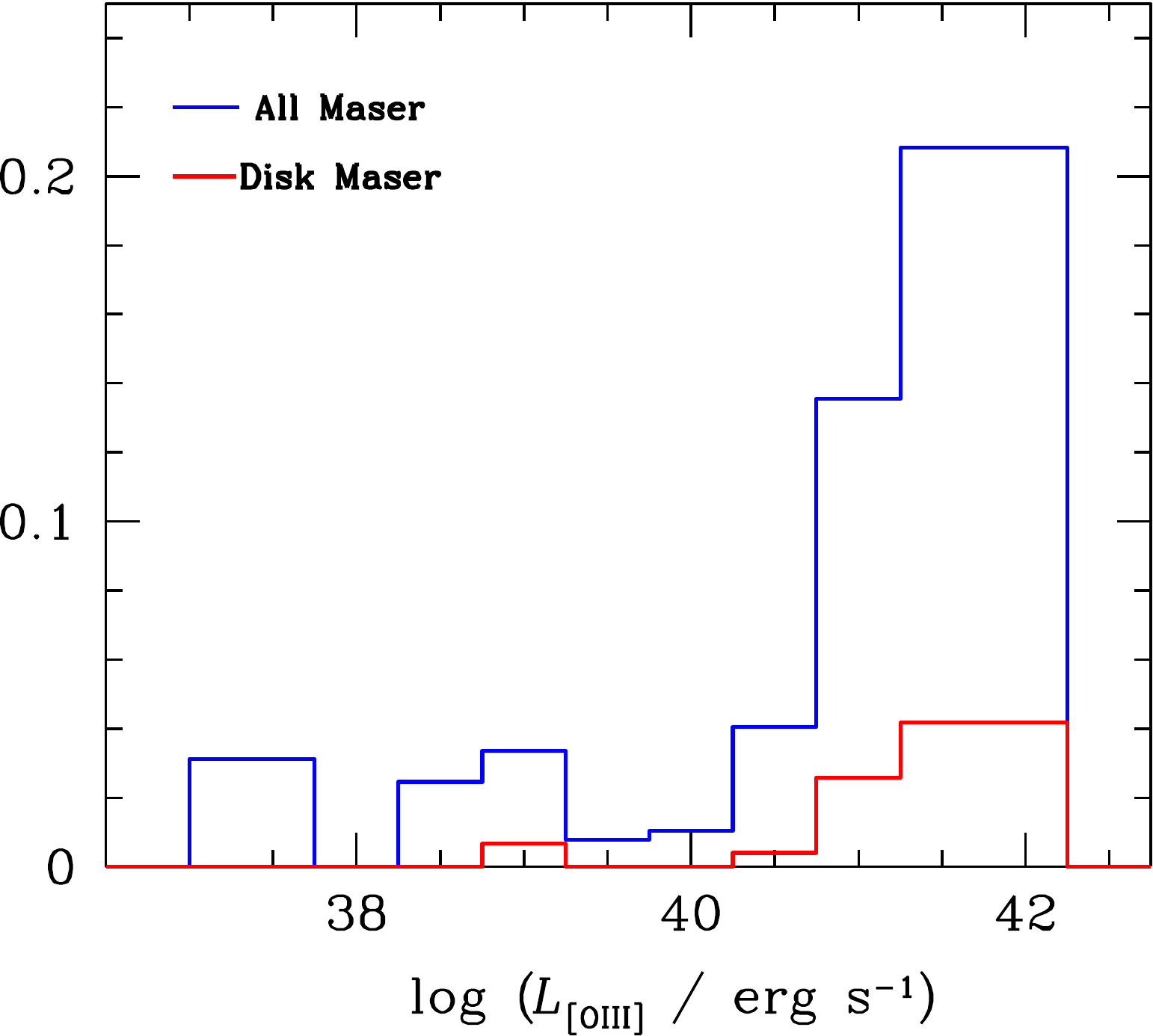}    } \vskip -1mm
	
\figcaption[]{ {\it Left}: We show the detection fraction as a function of stellar velocity dispersion. The fraction of masers of all sorts in the MCP sample (blue dashed) is compared to the detection fraction when the non-detections from this paper are included (solid blue). Likewise, we compare the maser disk detection fraction from Constantin et al. (red long-dashed line) with the corrected detection fraction when our sample is included (red solid).  These two lines are quite similar. In both panels, only bins with more than 30 objects are shown. It is clear that the apparent trend towards a higher detection fraction at higher dispersion is at least partially due to small number statistics in these bins. As we add more points, we see that the detection fractions do not significantly rise towards higher \sigmastar. In the highest dispersion bin the HETGBT more than doubles the number of searched objects. Note that there are only 30 masing disks and thus the maser detection rates suffer from low number statistics.  {\it Right}: Similar to above, we plot the maser detection fraction as a function of the [O {\tiny III}] luminosity. Again, only bins with at least 30 galaxies are shown. In this case the rising detection fraction towards more luminous active galaxies appears to be real. Again, only bins with at least 30 galaxies are shown. \label{fig:sigma_frac}} 
	
\end{center}
\end{figure*}

%%%%%%%%%%%%%%%%%%%%%%%%%%%%%%%%%%%%%%%%%%%%%%%%%%%%%%%%%%%%%%%%%%%%%%

\section{Detection Fractions} \label{sec:Results}

We did not detect any new masers in the 87 objects surveyed. The overall detection rate of masers is $<3$\% for all galaxies, hence our non-detections could just be due to low number statistics. Assuming an average detection rate of 3\% and using the simple binomial probability distribution, the probability of detecting no masers (7\%) is only \nicefrac{1}{3}\ of the probability of detecting 2 or 3 masers (25\% and 22\% respectively) given the sample size of 87 galaxies.

To put our non-detections into context, we show maser (and maser disk) detection fractions as a function of \sigmastar\ and $L_{\rm [OIII]}$ in Figure \ref{fig:sigma_frac}. Below we combine our sample with a subset of the galaxies searched with GBT that have optical spectroscopy from the literature.

\subsection{Nondetections}

As described above, the largest current megamaser disk search with uniform sensitivity has been carried out by the GBT. We combine the full list of galaxies searched by the GBT with our smaller list to search for trends between detections and optical properties of the galaxies. To maximize the number of searched galaxies with literature optical spectroscopy, we rely on the Sloan Digital Sky Survey. Constantin et al. (in prep) started with the 3339 maser non-detections as of June 2013, along with 151 galaxies with maser detections. They cross-matched these galaxies with the Sloan Digital Sky Survey \citep[SDSS][]{yorketal2000} and the Palomar Survey of nearby galaxies \citep{hfs1997spec}. There are spectroscopic matches for 1330 of the non-detections, and 92 of the maser galaxies, which include 15 maser disks. From these matches, we have measurements of stellar velocity dispersion (\sigmastar), and emission-line properties, including the Balmer decrement and the \oiii\ luminosity corrected for extinction. We will also discuss measurements of the maser luminosities, which typically fall in the range of tens to thousands of solar luminosities for megamasers. These represent the luminosity that the maser system would have if it were isotropic. The true luminosities are very uncertain as they are highly dependent on the (generally unknown) beaming distribution \citep{kuoetal2011}.

We should also note that there may well be subtle biases in the galaxies that have been targeted spectroscopically by the SDSS. These biases are then imprinted on the subset of targeted galaxies presented here. For instance, heavily reddened galaxies may be less likely to fall into the SDSS main galaxy sample \citep{straussetal2002}. More massive galaxies are also less likely to be targeted spectroscopically at very low redshift \citep[e.g.,][]{fukugitaetal2007}. Therefore, we should be wary of jumping to very strong conclusions until these biases are also studied and accounted for (Constantin et al. in preparation).

\subsection{Detection as a Function of Velocity Dispersion}

Both \citet{zhuetal2011} and \citet{constantin2012} pointed out a rising fraction of megamaser galaxies as the galaxy stellar velocity dispersion rises. However, there are very few galaxies in their samples with \sigmastar$> 160$~\kms. Nearly all of our galaxies fall in this high-dispersion regime. In Figure \ref{fig:sigma_frac} (left), we show the maser fraction as a function of \sigmastar\ prior to our survey (long-dashed lines) as well as the detection fraction after adding all of our non-detections (solid). Unfortunately, just based on Poisson errors alone, we still do not have enough data in the high-dispersion bins to make a significant measurement of the maser fraction at high dispersion. However, it is clear that the continued rise in detection fraction above \sigmastar$\approx 150$~\kms\ is not real. If we restrict our attention to the megamaser disk galaxies alone, then we still see a rise towards \sigmastar$\approx 150$~\kms, but the measurements at low dispersion are also highly uncertain due to small numbers.

\subsection{Detection as a Function of Luminosity}

Apart from the dependence on velocity dispersion, the trend seen most clearly in \citet{zhuetal2011} and \citet{constantin2012} is an increased detection fraction at higher \oiii\ luminosity. Recall that in Seyfert galaxies, the \oiii\ luminosity is an indirect indicator of the bolometric luminosity of the AGN \citep[e.g.,][]{yee1980}, and is often used when the non-thermal continuum cannot be directly measured \citep[e.g.,][]{zakamskaetal2003,heckmanetal2004,liuetal2009}.

In Figure \ref{fig:sigma_frac} (right) we again show the maser and maser disk detection fractions for the sample of galaxies searched by the MCP. The trend towards higher detection fraction at higher \oiii\ luminosity is quite clear in this case, even excluding uncertain bins. Because we do not have accurate measurements of the \oiii\ luminosity for the HETMGS sample, we do not include them. The ten HETGBT galaxies with SDSS spectra would fall at the faint end of the distribution, with $L_{\rm [OIII]} \approx 5 \times 10^{37}$~erg~s$^{-1}$. From this figure alone, one might conclude that our non-detections are due entirely to the luminosity distribution of the sources we targeted. If megamaser disk luminosity and AGN bolometric luminosity are correlated \citep[e.g.,][]{henkeletal2005, kondratkoetal2006a} then perhaps we simply did not have the sensitivity to detect the possibly very faint masers around these very weak AGN.

\begin{figure}
\includegraphics[width=0.45
\textwidth]{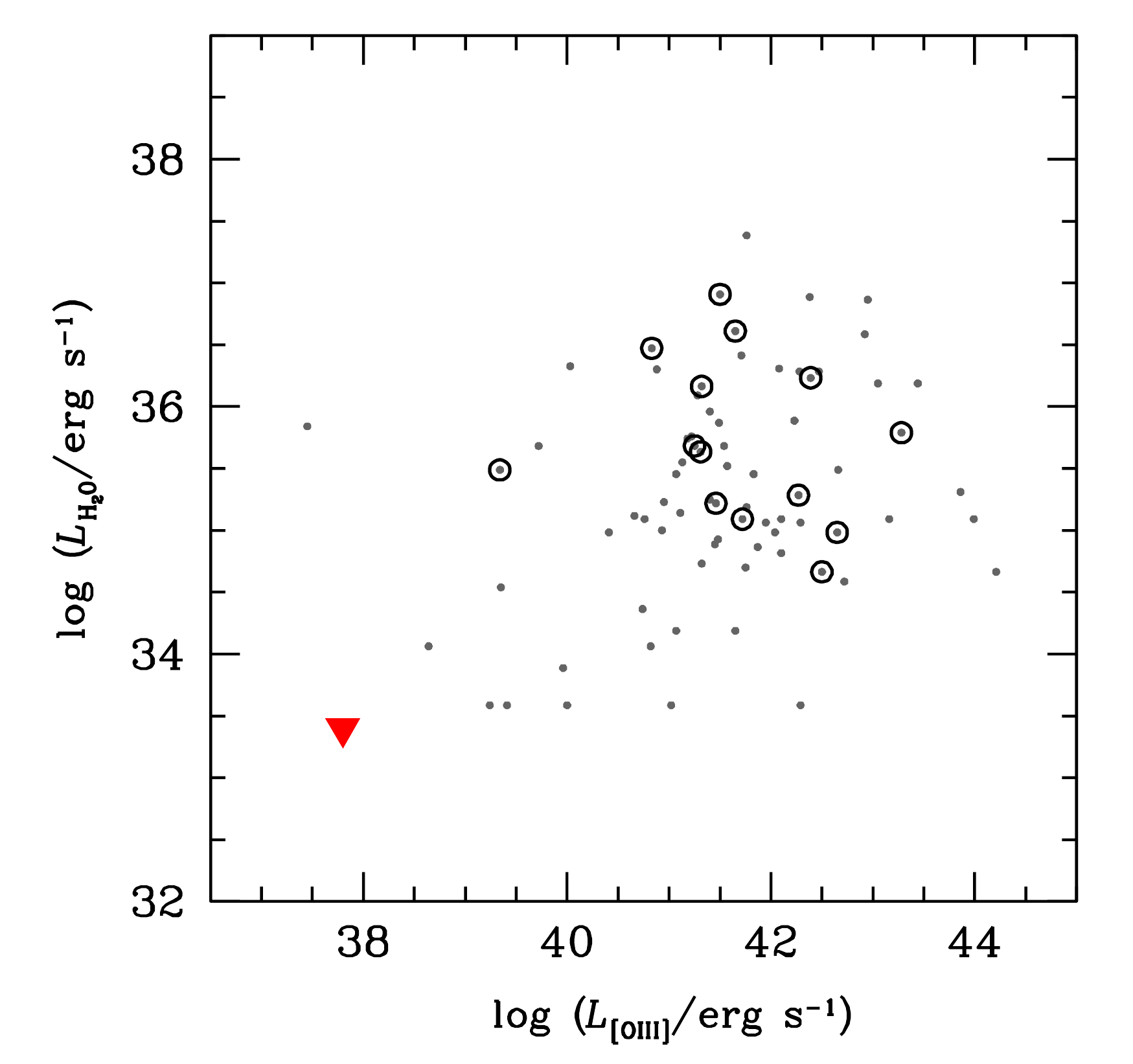}
\vskip -2mm
\figcaption[]{ Relation between 22 GHz H$_2$O maser emission luminosity (assumed isotropic) and the observed extinction-corrected [O {\tiny III}]$\lambda 5007$ luminosity for all known water megamasers in the SDSS (grey points). Megamaser disks are circled. The big red triangle is the $3\sigma$ upper limit on the maser luminosity of a source at the median sample distance of 74 Mpc, assuming our detection limit of 0.2 $L_{\odot}$. The [O {\tiny III}] luminosity is the median for the subsample of objects with SDSS spectra. The megamaser disk galaxies span a very narrow range in isotropic maser luminosity. This is expected, given the constant surface density predicted by \citet{neufeldetal1994} and the similar size ($\sim 0.5$ pc) of all the masing disks. \label{fig:flum}}
\end{figure}

However, we suggest that there is more to the story. Examination of Figure \ref{fig:flum} reveals that while there is a weak correlation between the maser luminosity and $L_{\rm [OIII]}$ when looking at all maser sources, this correlation vanishes for the maser disk sources taken alone. Instead, they span a very narrow range in isotropic maser luminosity that is virtually independent of $L_{\rm [OIII]}$. A similar trend can be seen when looking at a function of hard X-ray luminosity in \citet{kondratkoetal2006b}; most of the correlation is driven by the non-disk masers. At the same time, our typical $L_{\rm [OIII]}$ luminosity is well below the typical luminosity for known megamaser disks (including NGC4258). Perhaps we are detecting a true threshold in luminosity, below which the temperature condition for masing ($\sim 400$~K e.g., Neufeld et al. 1994) is not met. An $L_{\rm [OIII]} \approx 10^{38}$~erg~s$^{-1}$ corresponds roughly to $L_{\rm X} \approx 10^{40}-10^{41}$~erg~s$^{-1}$, depending on the bolometric correction and the assumed dust correction for $L_{\rm [OIII]}$ \citep{vasudevanfabian2007, liuetal2009, shaoetal13a}. This threshold, which lies below the observed X-ray luminosity of known maser disks \citep{kondratkoetal2006b}, is roughly consistent with the calculations of \citet{neufeldetal1994}. In \S 4, we discuss in more detail the physical connection between luminosity, BH mass, and disk size that may cause the dearth of masers we observe.

%%%%%%%%%%%%%%%%%%%%%%%%%%%%%%%%%%%%%%%%%%%%%%%%%%%%%%%%%%%%%%%%%%%%

%%%%%%%%%%%%%%%%%%%%%%%%%%%%%%%%%%%%%%%%%%%%%%%%%%%%%%%%%%%%%%%%%%%%%

\subsection{LINERs and Seyferts} \label{sec:liners}

We can also ask whether the detection fraction depends on the type of nuclear activity that we observe. In particular, while line ratios associated with high-ionization Seyfert galaxies are difficult to achieve through any mechanism other than accretion onto an SMBH, LINER activity has a wide array of causes, including shocks \citep[see review in ][]{ho2008}. Particularly in large-aperture data, previous studies \citep{sarzietal2006, singhetal13a} have shown that the central LINER emission often does not arise directly from an active galactic nucleus. Our HETGBT sample contains equal numbers of LINERs and Seyferts and almost all are low luminosity, as indicated by the low Amplitude-over-Noise of the emission lines \citep{vandenboschetal2015}.

The majority of the megamaser detections are found in Seyfert galaxies (see Figure \ref{fig:bpt}), likely because many of the LINERs in the SDSS are not actually powered by nuclear accretion. On the other hand, NGC 4258 is a LINER, so we had some hope that a survey focused on low-luminosity systems like NGC4258 would have a higher yield \citep{hfs1997broad}. That was not the case. We conclude that it is more important to select galaxies above the possible X-ray luminosity threshold than to worry directly about the optical line ratios.

\section{Discussion} \label{sec:Discussion}

Our detected maser fraction of $\lesssim 1$\% is nominally lower than that seen by the previous maser searches. While a larger sample is needed to be sure of this lower detection fraction, we here discuss in more detail the most likely causes of our low detection fraction and implications for future searches. The most obvious differences between this sample and previous samples are: (1) lower AGN luminosities and (2) more massive host galaxies with higher stellar velocity dispersions. As a result, the galaxies that we target likely have higher BH masses and very low mass accretion rates. Taking all of these properties into account, we investigate various explanations for the lack of maser detections. First we discuss the non-detections of megamaser disks, then we discuss the non-detections of megamasers of any sort.

\subsection{AGN Luminosity and Maser Disk Size}

It is easy to imagine that the megamaser luminosity will correlate with the luminosity of the AGN on average \citep{neufeld00a}. It is the X-ray corona that most likely heats the molecular accretion disk, thus creating a metastable population of excited water molecules that are able to mase \citep{herrnsteinetal2005}. A weak correlation is seen between hard X-ray luminosity and megamaser luminosity when all megamaser AGN are considered \citep{kondratkoetal2005}, as well as a correlation between megamaser luminosity and FIR luminosity \citep{henkeletal2005}. Furthermore, it is clear from Figure \ref{fig:sigma_frac} that the detection fraction of megamasers in general, and maser disks in specific, both rise at higher \oiii\ (and therefore bolometric) luminosity. However, as argued above, given the lack of correlation between megamaser disk luminosity and AGN luminosity (Fig. \ref{fig:flum}), we find it unlikely that our non-detections can be explained by the presence of very faint masers around our very low luminosity AGN. We favor the possibility that there is a threshold $L_{\rm X} \approx 10^{40}$~erg~s$^{-1}$, below which the conditions for masing are no longer met. Specifically, the size of the maser disk required at low accretion rate and high BH mass may simply grow too large to be stable.

According to \citet{neufeldetal1994}, masing disks are expected to have a very constant surface luminosity density. We thus expect the outer radius of the disk $R_{cr}$, set by the transition from a molecular to atomic disk, to scale with the X-ray luminosity \citep{kondratkoetal2006b}. Such a scaling is observed by \citet{wardle12}. Let us now imagine that we take NGC4258 as a model for the maser disks that we had hoped to find in our elliptical galaxy sample. NGC4258 is a good analog because both it and the ellipticals are accreting at very low fractions of their Eddington luminosity \citep{ho2008}. In fact, given the typical observed luminosities, and inferred BH masses of $\sim 10^8$~\msun, we do expect typical Eddington ratios of $\sim 10^{-4}$, as in NGC4258 \citep{herrnsteinetal2005}

We can imagine scaling the properties of the NGC4258 maser disk to infer the likely properties of comparable molecular disks around elliptical galaxies. If $R_{cr} \propto L_X$ \citep[as shown by][]{kondratkoetal2006b} then, provided that the Eddington fraction, accretion efficiency and the X-ray-to-bolometric efficiency stay roughly constant \citep[likely a valid assumption, e.g., ][]{vasudevanfabian2007}, we find that $R_{cr} \propto M_{\rm BH}$. That is, the size of the accretion disk would scale linearly with the BH mass. In our sample, we expect the BHs are roughly an order of magnitude more massive than the BH in NGC4258. Therefore, if this scaling roughly holds, their molecular disks would have sizes of $\sim 5-10$ pc.

This scaling between X-ray luminosity, BH mass, and Eddington ratio, introduces a possible explanation for the low incidence of megamaser disks in local elliptical galaxies. Because of the uniformly low Eddington ratio and high \mbh, the accretion disk size will naturally grow by at least an order of magnitude. Likewise the \citet{toomre64} $Q$ parameter would decrease by a full order of magnitude at constant disk surface density, as $Q$ depends linearly on the orbital frequency, and so inversely on $R$. It is not clear that accretion disks would be stable at a radius of 10 pc. Thus, one possible explanation for the lack of masers is that at high \mbh\ and low Eddington fraction, it is no longer physically possible to maintain a molecular disk with conditions appropriate for masing.

In actuality, the Eddington fraction in the ellipticals may be even lower than in NGC4258. In the Palomar spectroscopic survey of galaxies, \citet{ho2008} finds that LINERs radiate at typical Eddington fractions of $\sim 10^{-5}$, but span a large range from $10^{-7}$ to $10^{-3}$. If the Eddington ratio drops systematically towards more massive systems, the situation is more complicated than outlined in the previous paragraph and the disk size would not grow as rapidly with BH mass.

\subsection{Environmental Effects}

Another possibility is that the elliptical galaxy environment is less conducive to supporting molecular disks, even at a fixed Eddington fraction. Higher mass galaxies do host more hot X-ray gas, which could inhibit the formation of large amounts of dense molecular gas \citep{wiklind01,osullivan01,sarzietal2006}. Albeit on much larger scales, there are documented differences in the properties of molecular disks in elliptical and spiral galaxies. \citet{davisetal2014} find that the molecular disks in early-type galaxies have star-formation rates far below those expected from the Schmidt-Kennicutt relation \citep[e.g.,][]{kennicutt1998}. Their explanation is that the high levels of shear in the inner disk suppresses star formation. The maser disks are even deeper in the potential well, and we speculate that the gravitational potential may lead to a suppression of masing.

\subsection{Jet Masers}

Even if we do not detect any masing disks, given our sample size we might have expected to detect megamasers associated with jet activity. As is well-documented in the literature, jets grow more prevalent in massive elliptical galaxies \citep[e.g.,][]{matthewsetal1964,bestetal2005,mandelbaumetal2009} and a high fraction of elliptical galaxy nuclei contain low levels of radio emission \citep[e.g.,][]{sadleretal1989,wrobelheeschen1991}. Furthermore, elliptical galaxies with radio sources are also more likely to contain dust lanes \citep[e.g.,][]{vandokkumfranx1995} and dust is usually accompanied with molecular gas, which is needed to form or fuel the masing disk. Given the rising frequency of jet emission in elliptical galaxies, and particularly in the most luminous of elliptical galaxies, our non-detections are particularly interesting.

Here we might again invoke luminosity. Certainly jet power is known to correlate with \oiii\ luminosity \citep[e.g.,][]{hopeng2001}. Another culprit is the lower gas mass in the elliptical galaxy. It is possible that because there is less molecular gas along the path of the jet, there are fewer opportunities for masing. Another factor is the geometry of the jet, as the Doppler-boosted forward jet may lie in front of the bulk of the molecular gas \citep[see also][]{henkel98}.

\subsection{Megamaser Disks and Black Hole Demographics}

As mentioned above, one of our primary motivations in embarking on this survey was to study the reliability of stellar dynamical BH masses \citep[e.g.,][]{gebhardtetal2003} via comparison with the BH mass derived from a megamaser disk. Precious few galaxies today have BH mass measurements based on multiple techniques \citep[e.g.,][]{siopisetal2009, walshetal2013}, and even one such comparison in a massive elliptical would be extremely important.

Megamaser disks provide some of the only robust and unbiased measurements of \mbh\ in spiral galaxies. We have thus used these systems to measure scaling relations in late-type spiral galaxies that can be studied in detail with stellar dynamics. We see no compelling correlations between BH mass and galaxy properties in this regime \citep[e.g.,][]{greeneetal2010, sunetal2013}.

Thus far, known megamaser disks are found orbiting active galactic nuclei (AGNs) with BH masses that are clustered around $\sim 10^7$~\msun\ \citep{herrnsteinetal2005,kuoetal2011} and mostly found in massive spiral galaxies \citep{greeneetal2010}. This narrow range in properties is likely a natural biproduct of the selection technique. In general, megamaser searches have focused on known active galaxies \citep[e.g.,][]{braatzetal1997,kondratkoetal2006a,greenhilletal2009}, predominantly selected based on their location in optical diagnostic diagrams \citep{baldwinetal1981,veilleuxosterbrock1987}. Since the typical obscured active galaxy in optically selected samples has \mbh$\sim 10^7$~\msun\ \citep{heckmanetal2004,greeneho2007b}, it is not surprising that the majority of the discovered megamaser disk galaxies also have similar BH masses. At higher BH masses, the BHs are no longer highly active and thus are not included in the Seyfert galaxy catalogs. At lower BH masses, the dearth of sources is a selection effect due to (a) the difficulties of isolating accretion signatures in the presence of dust and star formation and (b) the fact that low-mass BHs are faint even when radiating at their Eddington limit \citep[e.g.,][]{greeneho2007b,reinesetal2013}.

Having failed to find a megamaser disk around an \mbh$>10^7$~\msun\ SMBH, we wish here to make a slightly different point about the potential biases inherent in stellar dynamical BH mass measurements \citep[e.g.,][]{gultekinetal2011}. In principle megamaser disks in massive elliptical galaxies have the potential to reveal a population of BHs that cannot be found using stellar dynamics. If the BH mass to galaxy mass ratio is small enough, then there will be no ``gravitational sphere of influence'' where the BH mass dominates the stellar dynamics. There will be no kinematic signature of the BH on the stars. On the other hand, a sub-pc scale megamaser disk could still be within the sphere of influence of the BH, and thus probe the (hitherto unexplored) regime of very low \mbh/$M_{\rm gal}$.

In closing, we re-emphasize the importance of megamaser disks in revealing BH demographics at low mass. At low $\sigma_\ast$ in particular, there are not many remaining galaxies in the universe where we should be able to resolve the gravitational sphere of influence with present-day technology \citep[assuming that $\sigma_\ast$ is a valid proxy;][ their Fig.~10]{batcheldor10a,gultekinetal2011,vandenboschetal2015}, and at larger dispersion and luminosities, the situation is only a little better. Even with new facilities like JWST, ALMA \citep{davis14}, and ELTs \citep{do14}, the improvement in spatial resolution will only probe one order of magnitude smaller in BH mass at a fixed galaxy property, which will not allow us to probe the full range in BH mass at $\sigma_\ast < 200$ \kms. Hence, even in the coming decades, masing disks and indirect AGN methods will remain critical for finding low-mass BHs \citep{reines15}.

\section{Conclusions} \label{sec:summary}

Using the GBT we have surveyed 87 galaxies with large dispersion ($>250$ \kms) in search of water megamaser emission at 22 GHz associated with AGN activity. The overall detection rate in previous surveys, which cover mostly spiral galaxies, is ~3\%, while here we detected no masers in our mostly elliptical sample. We discuss various explanations, including low number statistics, local changes related to \mbh\ and accretion rate leading to very large unstable disks, or environmental differences due to the deep potential or gas-free nature of elliptical galaxies.

The goal of our search was to find a masing disk in a galaxy in which dual BH mass measurement was possible to cross-calibrate different techniques. Continuing this search is worthwhile, but the yield could be low. Since taking the GBT observations reported here, more objects were added to the HETMGS, including 35 masers as special targets \citep[\S 2.5 in][]{vandenboschetal2015}. In total 480 of the 1022 galaxies in the HETMGS have been searched for masers. However most of the remaining HETMGS galaxies are too far away for a stellar dynamical black hole mass, or have no AGN-like emission lines. There remain only a dozen objects in the final HETMGS that satisfy the selection criteria of this work and have yet to be searched. Galaxies with large spheres-of-influence are inherently rare. Dropping the AGN-like emission-line requirement increases the remaining candidates to 350 objects that have not yet been searched for a maser, depending on the choice of black hole scaling relation. Given the value of such a maser for cross calibration of black hole mass measurements, this would be a worthwhile search.

Alternatively, the detection rate can be maximized by targetting galaxies with similar properties as existing maser disks. The SDSS-DR7 spectroscopic galaxy sample is fairly close to a random sampling of galaxies \citep{yorketal2000} and it contains -- hopefully representative -- 14 disk masers, all of which are within 156 Mpc, have \mbox{0.5 < log [O{\tiny III}]/H$_\beta$ < 1.2}, -0.5 < log [N{\tiny II}]/H$_\alpha$ < 0.5, $90 < \sigma_\ast < 190$ \kms\ and have intrinsic $L_{\rm [O{\tiny III}]} \approx 10^{39}-10^{43}$~erg~s$^{-1}$ (corrected based on Balmer decrements and a nominal $\lambda^{-0.7}$ extinction law). This is a very narrow window of galaxy properties. There are only 65 galaxies inside this box of properties in the SDSS spectroscopic sample that have already been searched unsuccessfully with the GBT for masers. Taking this number at face value provides a very high, a posteriori, detection rate of 18\%, given the 14 that were detected. If all 14+65=79 of these galaxies have a masing disk, this detection fraction implies an opening angle of 10 degrees, consistent with the 8 degrees observed in NGC4258 \citep{bragg00,herrnsteinetal2005}. Theoretical estimates of the opening angle are much smaller \citep[$0.1$ degrees,][]{lo2005}, however the observed beaming is increased by the warping that is often observed in these disks. There remain less than 430 spectra in the SDSS with the same properties, that have yet to be searched. At the same detection rate, we would find 34 more. However, we caution that the box we drew in physical properties was rather arbitrary and may not in fact increase our chances to observe additional masers. Furthermore, these new objects would not have large spheres of influence, nor would they alleviate the biases in the maser searches.

% selection box:
% intrinsic \oiii\  luminosities of  \approx 10^{40}-10^{41}$~erg~s$^{-1}$
%log L[O III]-intrinsic = 39 - 43 (corrected based on Balmer decrements and nominal lambda^(-0.7) extinction law)
%-they are all within 156 kMpc  luminosity distance 
%-range in logL[O III]-observed = 38 - 42
%-range in logL[O III]-intirnsic = 39 - 43 (corrected based on Balmer decrements and nominal lambda^(-0.7) extinction law)
%-range in sigma = 90 - 190 km/s
%-range in log O[III]/Hbeta = 0.5 - 1.2 —will go with > 0.5, as you recall.

Ultimately we need better statistics to determine definitively why elliptical galaxies may host fewer megamasers. It would be interesting to target samples with known radio emission, to see if we can find masers associated with jet activity, although previous searches done this way have not had a high yield \citep{henkel98}. Likewise, it would be useful to target sources with known gas disks on $\sim100$pc scales \citep[e.g.,][]{laueretal2005,martinietal2013} to see if these may be more likely to host masers. Many of these searches will require larger volumes than searched here, but recently megamasers disks have been detected with distances as large as 156 Mpc and so there is hope.

\acknowledgements 

\lettrine[slope=-2pt,nindent=-2pt,lines=2]{W}{e} want to thank Martin Bureau and Tim Davis for organising the SMBH Workshop in Oxford in March 2015, without which this manuscript would not have been resurrected. 

\noindent JEG acknowledges support from NSF grant AST-1310405.

\begin{table*}
\begin{center}

\relsize{-2}
\begin{tabular}{lccccccc}
\hline

Source &     RA &   Dec  & Vel &   Date  & T$_{sys}$ & Int. &  RMS \\
 &     &    & \kms\ &     & Kelvin & min.  &  mJy  \\

\hline \hline

NGC0050    & 00:14:44.60  & -07:20:42.0 & 5701  & 2012-11-09 & 45.8 &  10 &  4.6 \\ % & 4307  &   7107  \\
NGC0093    & 00:22:03.20  & +22:24:29.0 & 5380  & 2012-11-08 & 42.4 &  10 &  4.4 \\ % & 3989  &   6783  \\
NGC0311    & 00:57:32.70  & +30:16:51.0 & 5065  & 2012-11-08 & 44.5 &  10 &  4.8 \\ % & 3677  &   6465  \\
NGC0384    & 01:07:25.00  & +32:17:34.0 & 4233  & 2012-11-16 & 38.3 &  10 &  3.9 \\ % & 2853  &   5626  \\
NGC0430    & 01:12:59.90  & -00:15:09.1 & 5299  & 2012-11-16 & 44.6 &  10 &  4.6 \\ % & 3909  &   6702  \\
NGC0533    & 01:25:31.42  & +01:45:34.3 & 5549  & 2012-11-09 & 45.6 &   9 &  5.0 \\ % & 4157  &   6954  \\
NGC0550    & 01:26:42.60  & +02:01:20.9 & 5829  & 2012-11-09 & 46.2 &  10 &  5.1 \\ % & 4434  &   7236  \\
NGC0584    & 01:31:20.81  & -06:52:05.0 & 1802  & 2012-11-16 & 48.3 &  10 &  5.2 \\ % &  443  &   3173  \\
PGC006116  & 01:39:09.01  & +48:16:56.9 & 5467  & 2012-11-06 & 32.9 &  10 &  3.5 \\ % & 4075  &   6871  \\
NGC0898    & 02:23:20.40  & +41:57:05.1 & 5495  & 2012-11-06 & 33.0 &  10 &  3.5 \\ % & 4103  &   6899  \\
UGC01859   & 02:24:44.40  & +42:37:22.9 & 5917  & 2012-11-06 & 32.6 &  10 &  3.2 \\ % & 4521  &   7325  \\
NGC0982    & 02:35:24.89  & +40:52:11.0 & 5737  & 2012-11-16 & 37.1 &  10 &  3.9 \\ % & 4343  &   7144  \\
ARK090     & 02:42:29.00  & +18:09:53.0 & 9508  & 2012-11-09 & 44.5 &  10 &  4.7 \\ % & 8079  &  10950  \\
UGC02261   & 02:48:17.42  & +50:48:00.8 & 4903  & 2012-11-06 & 32.5 &  10 &  3.2 \\ % & 3517  &   6302  \\
NGC1153    & 02:58:10.30  & +03:21:43.0 & 3126  & 2012-11-16 & 41.6 &  10 &  4.2 \\ % & 1756  &   4509  \\
UGC02495   & 03:02:06.69  & +41:35:37.1 & 9135  & 2012-11-16 & 36.4 &  10 &  3.7 \\ % & 7710  &  10573  \\
NGC1208    & 03:06:11.90  & -09:32:29.1 & 4356  & 2012-11-16 & 44.7 &  10 &  4.7 \\ % & 2974  &   5750  \\
PGC138608  & 03:13:53.23  & +62:32:58.9 & 3050  & 2012-11-10 & 87.3 &   9 &  9.6 \\ % & 1680  &   4432  \\
UGC02755   & 03:29:23.91  & +39:47:32.0 & 7326  & 2012-11-07 & 45.4 &  10 &  4.9 \\ % & 5918  &   8747  \\
UGC02866   & 03:50:14.89  & +70:05:41.0 & 1232  & 2012-11-06 & 34.8 &  10 &  3.4 \\ % & -120  &   2597  \\
UGC02881   & 03:52:16.91  & +36:14:12.9 & 5764  & 2012-11-10 & 87.7 &  10 &  9.8 \\ % & 4370  &   7171  \\
NGC1465    & 03:53:31.90  & +32:29:34.0 & 4194  & 2012-11-16 & 38.0 &  10 &  3.9 \\ % & 2814  &   5586  \\
NGC1469    & 04:00:27.71  & +68:34:40.0 & 1102  & 2012-11-06 & 34.6 &  10 &  3.8 \\ % & -249  &   2466  \\
IC0356     & 04:07:46.90  & +69:48:45.1 &  895  & 2012-11-06 & 34.4 &  10 &  3.6 \\ % & -454  &   2257  \\
IC0359     & 04:12:28.29  & +27:42:07.1 & 4053  & 2012-11-07 & 44.9 &   9 &  4.9 \\ % & 2674  &   5444  \\
UGC03024   & 04:22:26.61  & +27:17:51.6 & 5236  & 2012-11-10 & 79.1 &  10 &  8.8 \\ % & 3847  &   6638  \\
2M04310    & 04:31:05.21  & +23:24:08.0 & 5105  & 2012-11-07 & 43.5 &  10 &  4.7 \\ % & 3717  &   6506  \\
PGC165398  & 04:31:57.09  & +59:25:47.0 & 4630  & 2012-11-07 & 40.0 &  10 &  4.1 \\ % & 3246  &   6026  \\
NGC1653    & 04:45:47.40  & -02:23:34.0 & 4331  & 2012-11-16 & 40.6 &  10 &  4.2 \\ % & 2950  &   5725  \\
UGC03386   & 06:02:37.89  & +65:22:16.0 & 4607  & 2012-11-07 & 37.1 &  10 &  3.7 \\ % & 3223  &   6003  \\
PGC019864  & 06:55:27.70  & +33:16:50.0 & 5302  & 2012-11-07 & 36.3 &  10 &  4.0 \\ % & 3912  &   6704  \\
PGC020827  & 07:22:10.90  & -05:55:47.1 & 1618  & 2012-11-08 & 43.1 &  10 &  4.7 \\ % &  261  &   2987  \\
UGC03855   & 07:28:13.04  & +58:30:23.8 & 3167  & 2012-11-07 & 35.2 &  10 &  3.5 \\ % & 1796  &   4550  \\
NGC2411    & 07:34:36.39  & +18:16:52.9 & 5073  & 2012-11-08 & 38.2 &  10 &  3.9 \\ % & 3685  &   6473  \\
NGC2522    & 08:06:13.52  & +17:42:23.1 & 4705  & 2012-11-07 & 37.4 &  10 &  3.9 \\ % & 3320  &   6102  \\
PGC023680  & 08:26:24.91  & +59:53:42.8 & 7993  & 2012-11-07 & 33.1 &  10 &  3.3 \\ % & 6578  &   9420  \\
MRK1216    & 08:28:47.10  & -06:56:25.1 & 6394  & 2012-11-08 & 40.1 &  10 &  4.4 \\ % & 4994  &   7807  \\
NGC2787    & 09:19:18.49  & +69:12:12.1 &  696  & 2012-11-06 & 36.8 &  10 &  3.8 \\ % & -651  &   2056  \\
NGC3277    & 10:32:55.50  & +28:30:42.1 & 1408  & 2012-11-08 & 36.6 &  10 &  3.8 \\ % &   53  &   2775  \\
IC0624     & 10:36:15.19  & -08:20:02.1 & 5042  & 2012-11-08 & 40.2 &  10 &  4.1 \\ % & 3654  &   6442  \\
NGC3348    & 10:47:09.98  & +72:50:23.1 & 2837  & 2012-11-10 & 64.3 &  10 &  6.1 \\ % & 1469  &   4217  \\
PGC032873  & 10:56:15.99  & +42:19:58.9 & 7471  & 2012-11-10 & 56.3 &  10 &  5.7 \\ % & 6061  &   8893  \\
PGC036650  & 11:45:27.69  & +20:48:26.1 & 6935  & 2012-11-08 & 36.6 &  10 &  3.9 \\ % & 5530  &   8352  \\
NGC3869    & 11:45:45.61  & +10:49:29.1 & 3043  & 2012-11-08 & 38.7 &  10 &  3.9 \\ % & 1674  &   4425  \\
NGC3894    & 11:48:50.42  & +59:24:56.0 & 3223  & 2012-11-06 & 41.6 &  10 &  4.8 \\ % & 1852  &   4606  \\
NGC3919    & 11:50:41.51  & +20:00:53.9 & 6195  & 2012-11-14 & 37.5 &  10 &  4.1 \\ % & 4797  &   7606  \\
NGC3992    & 11:57:36.01  & +53:22:28.0 & 1048  & 2012-11-10 & 61.4 &  10 &  5.9 \\ % & -303  &   2411  \\
NGC4125    & 12:08:06.00  & +65:10:27.1 & 1356  & 2012-11-06 & 41.8 &  10 &  4.6 \\ % &    2  &   2722  \\
NGC4256    & 12:18:43.01  & +65:53:53.2 & 2528  & 2012-11-06 & 43.0 &  10 &  4.8 \\ % & 1163  &   3905  \\
NGC4403    & 12:26:12.81  & -07:41:06.0 & 5200  & 2012-11-10 & 67.8 &  10 &  7.7 \\ % & 3811  &   6601  \\
NGC4646    & 12:42:52.19  & +54:51:22.0 & 4647  & 2012-11-06 & 45.2 &  10 &  4.9 \\ % & 3263  &   6043  \\
NGC4673    & 12:45:34.70  & +27:03:39.3 & 6852  & 2012-11-10 & 59.5 &  10 &  5.9 \\ % & 5448  &   8269  \\
NGC4786    & 12:54:32.42  & -06:51:34.1 & 4647  & 2012-11-14 & 45.8 &   5 &  6.6 \\ % & 3263  &   6043  \\
NGC4958    & 13:05:48.90  & -08:01:13.0 & 1455  & 2012-11-10 & 71.7 &   9 &  8.0 \\ % &  100  &   2822  \\
PGC1021091 & 13:09:26.99  & -07:18:45.0 & 6723  & 2012-11-10 & 67.4 &  10 &  6.9 \\ % & 5320  &   8139  \\
NGC5133    & 13:24:52.90  & -04:04:55.1 & 6132  & 2012-11-10 & 68.3 &  10 &  6.9 \\ % & 4734  &   7542  \\
NGC5228    & 13:34:35.09  & +34:46:41.0 & 7706  & 2012-11-10 & 60.9 &  10 &  6.2 \\ % & 6294  &   9131  \\
IC0948     & 13:52:26.69  & +14:05:28.1 & 6912  & 2012-11-14 & 35.8 &  10 &  3.2 \\ % & 5507  &   8329  \\
NGC5400    & 14:00:37.23  & -02:51:28.1 & 7437  & 2012-11-14 & 37.7 &  10 &  4.3 \\ % & 6028  &   8859  \\
NGC5463    & 14:06:10.50  & +09:21:12.1 & 7178  & 2012-11-14 & 35.7 &  10 &  3.8 \\ % & 5771  &   8598  \\
NGC5623    & 14:27:08.71  & +33:15:07.0 & 3356  & 2012-11-14 & 35.9 &  10 &  3.6 \\ % & 1984  &   4740  \\
NGC5739    & 14:42:28.89  & +41:50:32.1 & 5377  & 2012-11-10 & 65.7 &  10 &  6.2 \\ % & 3986  &   6780  \\
UGC09602   & 14:55:55.20  & +11:51:41.0 & 9652  & 2012-11-08 & 45.1 &  10 &  4.7 \\ % & 8222  &  11095  \\
UGC09937   & 15:37:22.92  & +20:32:58.7 & 4526  & 2012-11-08 & 43.1 &  10 &  4.2 \\ % & 3143  &   5921  \\
UGC10097   & 15:55:43.30  & +47:52:01.9 & 5962  & 2012-11-10 & 70.7 &  10 &  7.3 \\ % & 4566  &   7370  \\
IC1153     & 15:57:03.02  & +48:10:05.9 & 5919  & 2012-11-10 & 69.2 &  10 &  7.2 \\ % & 4523  &   7327  \\
NGC6036    & 16:04:30.69  & +03:52:07.1 & 5505  & 2012-11-08 & 44.9 &  10 &  4.6 \\ % & 4113  &   6909  \\
NGC6146    & 16:25:10.31  & +40:53:34.0 & 8820  & 2012-11-10 & 72.0 &  10 &  7.0 \\ % & 7398  &  10255  \\
NGC6240    & 16:52:58.90  & +02:24:03.0 & 7465  & 2012-11-08 & 43.1 &   9 &  5.0 \\ % & 6055  &   8887  \\
PGC1347752 & 17:36:11.10  & +08:28:56.0 &  814  & 2012-11-08 & 41.4 &  10 &  4.4 \\ % & -535  &   2175  \\
NGC6508    & 17:49:46.47  & +72:01:16.0 & 7637  & 2012-11-10 & 75.1 &  10 &  7.3 \\ % & 6226  &   9061  \\
UGC11082   & 18:00:05.50  & +26:22:00.0 & 4739  & 2012-11-08 & 39.5 &  10 &  4.0 \\ % & 3354  &   6136  \\
NGC6548    & 18:05:59.20  & +18:35:14.0 & 2209  & 2012-11-08 & 41.0 &  20 &  3.1 \\ % &  847  &   3583  \\
NGC6619    & 18:18:55.51  & +23:39:20.1 & 5038  & 2012-11-14 & 35.2 &  10 &  3.4 \\ % & 3650  &   6438  \\
PGC062122  & 18:36:39.70  & +19:43:45.0 & 4840  & 2012-11-14 & 35.8 &  10 &  3.5 \\ % & 3454  &   6238  \\
NGC6688    & 18:40:40.11  & +36:17:23.0 & 5462  & 2012-11-08 & 38.5 &  10 &  4.1 \\ % & 4071  &   6866  \\
UGC11353   & 18:47:44.20  & +23:20:49.9 & 4208  & 2012-11-14 & 35.8 &  10 &  3.6 \\ % & 2828  &   5600  \\
NGC6921    & 20:28:28.80  & +25:43:23.9 & 4337  & 2012-11-08 & 39.0 &  10 &  4.0 \\ % & 2956  &   5731  \\
PGC066592  & 21:20:42.50  & +44:23:58.9 & 3894  & 2012-11-16 & 48.5 &  10 &  5.5 \\ % & 2517  &   5284  \\
UGC11920   & 22:08:27.40  & +48:26:27.1 & 1103  & 2012-11-16 & 44.5 &  10 &  4.9 \\ % & -248  &   2467  \\
NGC7391    & 22:50:36.10  & -01:32:41.0 & 3048  & 2012-11-09 & 43.5 &  10 &  4.5 \\ % & 1678  &   4430  \\
NGC7426    & 22:56:02.80  & +36:21:40.9 & 5325  & 2012-11-08 & 39.6 &  10 &  4.3 \\ % & 3935  &   6728  \\
NGC7436    & 22:57:57.50  & +26:09:00.0 & 7375  & 2012-11-08 & 38.8 &  10 &  4.1 \\ % & 5966  &   8797  \\
IC5285     & 23:06:58.90  & +22:56:10.9 & 6154  & 2012-11-16 & 43.6 &  10 &  4.8 \\ % & 4756  &   7564  \\
NGC7671    & 23:27:19.30  & +12:28:03.0 & 4128  & 2012-11-08 & 41.0 &  10 &  4.4 \\ % & 2749  &   5520  \\
NGC7728    & 23:40:00.80  & +27:08:01.0 & 9398  & 2012-11-08 & 39.5 &  10 &  3.9 \\ % & 7970  &  10839  \\
\hline
\hline                          
\end{tabular}
\relsize{+2}
  \caption{ \textnormal{List of the observed 87 HETGBT galaxies observed with the GBT to search for megamasers. Column (1) HETMGS name,  (2,3) J2000 position, (4) optical LSRK velocity used for tuning the spectral windows, (5) Observation date, (6) System temperature in Kelvin, (7) Integration time in minutes, (8) sensitivity in mJy.}\label{tabledata}
}

\end{center}

\end{table*}


\begin{thebibliography}{}
\providecommand\natexlab[1]{#1}
\providecommand\JournalTitle[1]{#1}

\bibitem[{{Baldwin} {et~al.}(1981){Baldwin}, {Phillips}, \&
  {Terlevich}}]{baldwinetal1981}
{Baldwin}, J.~A., {Phillips}, M.~M., \& {Terlevich}, R. 1981,
  \href{http://adsabs.harvard.edu/abs/1981PASP...93....5B}{\JournalTitle{\pasp},
  93, 5}

\bibitem[{{Batcheldor}(2010)}]{batcheldor10a}
{Batcheldor}, D. 2010,
  \href{http://adsabs.harvard.edu/abs/2010ApJ...711L.108B}{\JournalTitle{\apjl},
  711, L108}

\bibitem[{{Beifiori} {et~al.}(2012){Beifiori}, {Courteau}, {Corsini}, \&
  {Zhu}}]{beifiori12}
{Beifiori}, A., {Courteau}, S., {Corsini}, E.~M., \& {Zhu}, Y. 2012,
  \href{http://adsabs.harvard.edu/abs/2012MNRAS.419.2497B}{\JournalTitle{\mnras},
  419, 2497}

\bibitem[{{Best} {et~al.}(2005){Best}, {Kauffmann}, {Heckman}, {Brinchmann},
  {Charlot}, {Ivezi{\'c}}, \& {White}}]{bestetal2005}
{Best}, P.~N., {Kauffmann}, G., {Heckman}, T.~M., {et~al.} 2005,
  \href{http://adsabs.harvard.edu/abs/2005MNRAS.362...25B}{\JournalTitle{\mnras},
  362, 25}

\bibitem[{{Blanton} \& {Moustakas}(2009)}]{blanton09}
{Blanton}, M.~R., \& {Moustakas}, J. 2009,
  \href{http://adsabs.harvard.edu/abs/2009ARA%26A..47..159B}{\JournalTitle{\araa},
  47, 159}

\bibitem[{{Braatz} {et~al.}(2010){Braatz}, {Reid}, {Humphreys}, {Henkel},
  {Condon}, \& {Lo}}]{braatzetal2010}
{Braatz}, J.~A., {Reid}, M.~J., {Humphreys}, E.~M.~L., {et~al.} 2010,
  \href{http://adsabs.harvard.edu/abs/2010ApJ...718..657B}{\JournalTitle{\apj},
  718, 657}

\bibitem[{{Braatz} {et~al.}(1994){Braatz}, {Wilson}, \&
  {Henkel}}]{braatzetal1994}
{Braatz}, J.~A., {Wilson}, A.~S., \& {Henkel}, C. 1994,
  \href{http://adsabs.harvard.edu/abs/1994ApJ...437L..99B}{\JournalTitle{\apjl},
  437, L99}

\bibitem[{{Braatz} {et~al.}(1997){Braatz}, {Wilson}, \&
  {Henkel}}]{braatzetal1997}
---. 1997,
  \href{http://adsabs.harvard.edu/abs/1997ApJS..110..321B}{\JournalTitle{\apjs},
  110, 321}

\bibitem[{{Bragg} {et~al.}(2000){Bragg}, {Greenhill}, {Moran}, \&
  {Henkel}}]{bragg00}
{Bragg}, A.~E., {Greenhill}, L.~J., {Moran}, J.~M., \& {Henkel}, C. 2000,
  \href{http://adsabs.harvard.edu/abs/2000ApJ...535...73B}{\JournalTitle{\apj},
  535, 73}

\bibitem[{{Condon} {et~al.}(1998){Condon}, {Cotton}, {Greisen}, {Yin},
  {Perley}, {Taylor}, \& {Broderick}}]{condon98}
{Condon}, J.~J., {Cotton}, W.~D., {Greisen}, E.~W., {et~al.} 1998,
  \href{http://adsabs.harvard.edu/abs/1998AJ....115.1693C}{\JournalTitle{\aj},
  115, 1693}

\bibitem[{{Constantin}(2012)}]{constantin2012}
{Constantin}, A. 2012,
  \href{http://adsabs.harvard.edu/abs/2012JPhCS.372a2047C}{\JournalTitle{Journal
  of Physics Conference Series}, 372, 012047}

\bibitem[{{Davis}(2014)}]{davis14}
{Davis}, T.~A. 2014,
  \href{http://adsabs.harvard.edu/abs/2014MNRAS.443..911D}{\JournalTitle{\mnras},
  443, 911}

\bibitem[{{Davis} {et~al.}(2014){Davis}, {Young}, {Crocker}, {Bureau}, {Blitz},
  {Alatalo}, {Emsellem}, {Naab}, {Bayet}, {Bois}, {Bournaud}, {Cappellari},
  {Davies}, {de Zeeuw}, {Duc}, {Khochfar}, {Krajnovi{\'c}}, {Kuntschner},
  {McDermid}, {Morganti}, {Oosterloo}, {Sarzi}, {Scott}, {Serra}, \&
  {Weijmans}}]{davisetal2014}
{Davis}, T.~A., {Young}, L.~M., {Crocker}, A.~F., {et~al.} 2014,
  \href{http://adsabs.harvard.edu/abs/2014MNRAS.444.3427D}{\JournalTitle{\mnras},
  444, 3427}

\bibitem[{{Do} {et~al.}(2014){Do}, {Wright}, {Barth}, {Barton}, {Simard},
  {Larkin}, {Moore}, {Wang}, \& {Ellerbroek}}]{do14}
{Do}, T., {Wright}, S.~A., {Barth}, A.~J., {et~al.} 2014,
  \href{http://adsabs.harvard.edu/abs/2014AJ....147...93D}{\JournalTitle{\aj},
  147, 93}

\bibitem[{{Drehmer} {et~al.}(2015){Drehmer}, {Storchi-Bergmann}, {Ferrari},
  {Cappellari}, \& {Riffel}}]{drehmer15}
{Drehmer}, D.~A., {Storchi-Bergmann}, T., {Ferrari}, F., {Cappellari}, M., \&
  {Riffel}, R.~A. 2015,
  \href{http://adsabs.harvard.edu/abs/2015MNRAS.450..128A}{\JournalTitle{\mnras},
  450, 128}

\bibitem[{{Falcke} {et~al.}(2000){Falcke}, {Wilson}, {Henkel}, {Brunthaler}, \&
  {Braatz}}]{falcke00}
{Falcke}, H., {Wilson}, A.~S., {Henkel}, C., {Brunthaler}, A., \& {Braatz},
  J.~A. 2000,
  \href{http://adsabs.harvard.edu/abs/2000ApJ...530L..13F}{\JournalTitle{\apjl},
  530, L13}

\bibitem[{{Fukugita} {et~al.}(2007){Fukugita}, {Nakamura}, {Okamura}, {Yasuda},
  {Barentine}, {Brinkmann}, {Gunn}, {Harvanek}, {Ichikawa}, {Lupton},
  {Schneider}, {Strauss}, \& {York}}]{fukugitaetal2007}
{Fukugita}, M., {Nakamura}, O., {Okamura}, S., {et~al.} 2007,
  \href{http://adsabs.harvard.edu/abs/2007AJ....134..579F}{\JournalTitle{\aj},
  134, 579}

\bibitem[{{Gebhardt} {et~al.}(2003){Gebhardt}, {Richstone}, {Tremaine},
  {Lauer}, {Bender}, {Bower}, {Dressler}, {Faber}, {Filippenko}, {Green},
  {Grillmair}, {Ho}, {Kormendy}, {Magorrian}, \& {Pinkney}}]{gebhardtetal2003}
{Gebhardt}, K., {Richstone}, D., {Tremaine}, S., {et~al.} 2003,
  \href{http://adsabs.harvard.edu/abs/2003ApJ...583...92G}{\JournalTitle{\apj},
  583, 92}

\bibitem[{{Gebhardt} {et~al.}(2007){Gebhardt}, {Lauer}, {Pinkney}, {Bender},
  {Richstone}, {Aller}, {Bower}, {Dressler}, {Faber}, {Filippenko}, {Green},
  {Ho}, {Kormendy}, {Siopis}, \& {Tremaine}}]{gebhardtetal07}
{Gebhardt}, K., {Lauer}, T.~R., {Pinkney}, J., {et~al.} 2007,
  \href{http://adsabs.harvard.edu/abs/2007ApJ...671.1321G}{\JournalTitle{\apj},
  671, 1321}

\bibitem[{{Greene} \& {Ho}(2007)}]{greeneho2007b}
{Greene}, J.~E., \& {Ho}, L.~C. 2007,
  \href{http://adsabs.harvard.edu/abs/2007ApJ...667..131G}{\JournalTitle{\apj},
  667, 131}

\bibitem[{{Greene} {et~al.}(2010)}]{greeneetal2010}
{Greene}, J.~E., {et~al.} 2010,
  \href{http://adsabs.harvard.edu/abs/2010ApJ...721...26G}{\JournalTitle{\apj},
  721, 26}

\bibitem[{{Greenhill} {et~al.}(2009){Greenhill}, {Kondratko}, {Moran}, \&
  {Tilak}}]{greenhilletal2009}
{Greenhill}, L.~J., {Kondratko}, P.~T., {Moran}, J.~M., \& {Tilak}, A. 2009,
  \href{http://adsabs.harvard.edu/abs/2009ApJ...707..787G}{\JournalTitle{\apj},
  707, 787}

\bibitem[{{G{\"u}ltekin} {et~al.}(2011){G{\"u}ltekin}, {Tremaine}, {Loeb}, \&
  {Richstone}}]{gultekinetal2011}
{G{\"u}ltekin}, K., {Tremaine}, S., {Loeb}, A., \& {Richstone}, D.~O. 2011,
  \href{http://adsabs.harvard.edu/abs/2011ApJ...738...17G}{\JournalTitle{\apj},
  738, 17}

\bibitem[{{G{\"u}ltekin} {et~al.}(2009){G{\"u}ltekin}, {Richstone}, {Gebhardt},
  {Lauer}, {Tremaine}, {Aller}, {Bender}, {Dressler}, {Faber}, {Filippenko},
  {Green}, {Ho}, {Kormendy}, {Magorrian}, {Pinkney}, \&
  {Siopis}}]{gultekinetal2009}
{G{\"u}ltekin}, K., {Richstone}, D.~O., {Gebhardt}, K., {et~al.} 2009,
  \href{http://adsabs.harvard.edu/abs/2009ApJ...698..198G}{\JournalTitle{\apj},
  698, 198}

\bibitem[{{Heckman} {et~al.}(2004){Heckman}, {Kauffmann}, {Brinchmann},
  {Charlot}, {Tremonti}, \& {White}}]{heckmanetal2004}
{Heckman}, T.~M., {Kauffmann}, G., {Brinchmann}, J., {et~al.} 2004,
  \href{http://adsabs.harvard.edu/abs/2004ApJ...613..109H}{\JournalTitle{\apj},
  613, 109}

\bibitem[{{Heckman} {et~al.}(1981){Heckman}, {Miley}, {van Breugel}, \&
  {Butcher}}]{heckmanetal1981}
{Heckman}, T.~M., {Miley}, G.~K., {van Breugel}, W.~J.~M., \& {Butcher}, H.~R.
  1981,
  \href{http://adsabs.harvard.edu/abs/1981ApJ...247..403H}{\JournalTitle{\apj},
  247, 403}

\bibitem[{{Henkel} {et~al.}(2005){Henkel}, {Peck}, {Tarchi}, {Nagar}, {Braatz},
  {Castangia}, \& {Moscadelli}}]{henkeletal2005}
{Henkel}, C., {Peck}, A.~B., {Tarchi}, A., {et~al.} 2005,
  \href{http://adsabs.harvard.edu/abs/2005A%26A...436...75H}{\JournalTitle{\aap},
  436, 75}

\bibitem[{{Henkel} {et~al.}(1998){Henkel}, {Wang}, {Falcke}, {Wilson}, \&
  {Braatz}}]{henkel98}
{Henkel}, C., {Wang}, Y.~P., {Falcke}, H., {Wilson}, A.~S., \& {Braatz}, J.~A.
  1998,
  \href{http://adsabs.harvard.edu/abs/1998A%26A...335..463H}{\JournalTitle{\aap},
  335, 463}

\bibitem[{{Herrnstein} {et~al.}(2005){Herrnstein}, {Moran}, {Greenhill}, \&
  {Trotter}}]{herrnsteinetal2005}
{Herrnstein}, J.~R., {Moran}, J.~M., {Greenhill}, L.~J., \& {Trotter}, A.~S.
  2005,
  \href{http://adsabs.harvard.edu/abs/2005ApJ...629..719H}{\JournalTitle{\apj},
  629, 719}

\bibitem[{{Ho}(2008)}]{ho2008}
{Ho}, L.~C. 2008,
  \href{http://adsabs.harvard.edu/abs/2008ARA%26A..46..475H}{\JournalTitle{\araa},
  46, 475}

\bibitem[{{Ho} {et~al.}(1997{\natexlab{a}}){Ho}, {Filippenko}, \&
  {Sargent}}]{hfs1997spec}
{Ho}, L.~C., {Filippenko}, A.~V., \& {Sargent}, W.~L.~W. 1997{\natexlab{a}},
  \href{http://adsabs.harvard.edu/abs/1997ApJS..112..315H}{\JournalTitle{\apjs},
  112, 315}

\bibitem[{{Ho} {et~al.}(1997{\natexlab{b}}){Ho}, {Filippenko}, {Sargent}, \&
  {Peng}}]{hfs1997broad}
{Ho}, L.~C., {Filippenko}, A.~V., {Sargent}, W.~L.~W., \& {Peng}, C.~Y.
  1997{\natexlab{b}},
  \href{http://adsabs.harvard.edu/abs/1997ApJS..112..391H}{\JournalTitle{\apjs},
  112, 391}

\bibitem[{{Ho} \& {Peng}(2001)}]{hopeng2001}
{Ho}, L.~C., \& {Peng}, C.~Y. 2001,
  \href{http://adsabs.harvard.edu/abs/2001ApJ...555..650H}{\JournalTitle{\apj},
  555, 650}

\bibitem[{{Houghton} {et~al.}(2006){Houghton}, {Magorrian}, {Sarzi}, {Thatte},
  {Davies}, \& {Krajnovi{\'c}}}]{Houghtonetal06}
{Houghton}, R.~C.~W., {Magorrian}, J., {Sarzi}, M., {et~al.} 2006,
  \href{http://adsabs.harvard.edu/abs/2006MNRAS.367....2H}{\JournalTitle{\mnras},
  367, 2}

\bibitem[{{Humphreys} {et~al.}(2013){Humphreys}, {Reid}, {Moran}, {Greenhill},
  \& {Argon}}]{humphreysetal2013}
{Humphreys}, E.~M.~L., {Reid}, M.~J., {Moran}, J.~M., {Greenhill}, L.~J., \&
  {Argon}, A.~L. 2013,
  \href{http://adsabs.harvard.edu/abs/2013ApJ...775...13H}{\JournalTitle{\apj},
  775, 13}

\bibitem[{{Jahnke} \& {Macci{\`o}}(2011)}]{jahnkemaccio2011}
{Jahnke}, K., \& {Macci{\`o}}, A.~V. 2011,
  \href{http://adsabs.harvard.edu/abs/2011ApJ...734...92J}{\JournalTitle{\apj},
  734, 92}

\bibitem[{{Kennicutt}(1998)}]{kennicutt1998}
{Kennicutt}, Jr., R.~C. 1998,
  \href{http://adsabs.harvard.edu/abs/1998ARA%26A..36..189K}{\JournalTitle{\araa},
  36, 189}

\bibitem[{{Kewley} {et~al.}(2006){Kewley}, {Groves}, {Kauffmann}, \&
  {Heckman}}]{kewley06}
{Kewley}, L.~J., {Groves}, B., {Kauffmann}, G., \& {Heckman}, T. 2006,
  \href{http://adsabs.harvard.edu/abs/2006MNRAS.372..961K}{\JournalTitle{\mnras},
  372, 961}

\bibitem[{{Kondratko} {et~al.}(2005){Kondratko}, {Greenhill}, \&
  {Moran}}]{kondratkoetal2005}
{Kondratko}, P.~T., {Greenhill}, L.~J., \& {Moran}, J.~M. 2005,
  \href{http://adsabs.harvard.edu/abs/2005ApJ...618..618K}{\JournalTitle{\apj},
  618, 618}

\bibitem[{{Kondratko} {et~al.}(2006{\natexlab{a}}){Kondratko}, {Greenhill}, \&
  {Moran}}]{kondratkoetal2006b}
---. 2006{\natexlab{a}},
  \href{http://adsabs.harvard.edu/abs/2006ApJ...652..136K}{\JournalTitle{\apj},
  652, 136}

\bibitem[{{Kondratko} {et~al.}(2006{\natexlab{b}}){Kondratko}, {Greenhill},
  {Moran}, {Lovell}, {Kuiper}, {Jauncey}, {Cameron}, {G{\'o}mez},
  {Garc{\'{\i}}a-Mir{\'o}}, {Moll}, {de Gregorio-Monsalvo}, \&
  {Jim{\'e}nez-Bail{\'o}n}}]{kondratkoetal2006a}
{Kondratko}, P.~T., {Greenhill}, L.~J., {Moran}, J.~M., {et~al.}
  2006{\natexlab{b}},
  \href{http://adsabs.harvard.edu/abs/2006ApJ...638..100K}{\JournalTitle{\apj},
  638, 100}

\bibitem[{{Kormendy} \& {Ho}(2013)}]{kormendyho2013}
{Kormendy}, J., \& {Ho}, L.~C. 2013,
  \href{http://adsabs.harvard.edu/abs/2013ARA%26A..51..511K}{\JournalTitle{\araa},
  51, 511}

\bibitem[{{Kuo} {et~al.}(2013){Kuo}, {Braatz}, {Reid}, {Lo}, {Condon},
  {Impellizzeri}, \& {Henkel}}]{kuoetal2013}
{Kuo}, C.~Y., {Braatz}, J.~A., {Reid}, M.~J., {et~al.} 2013,
  \href{http://adsabs.harvard.edu/abs/2013ApJ...767..155K}{\JournalTitle{\apj},
  767, 155}

\bibitem[{{Kuo} {et~al.}(2011)}]{kuoetal2011}
{Kuo}, C.~Y., {et~al.} 2011,
  \href{http://adsabs.harvard.edu/abs/2011ApJ...727...20K}{\JournalTitle{\apj},
  727, 20}

\bibitem[{{Kuo} {et~al.}(2015){Kuo}, {Braatz}, {Lo}, {Reid}, {Suyu}, {Pesce},
  {Condon}, {Henkel}, \& {Impellizzeri}}]{kuoetal2015}
{Kuo}, C.~Y., {Braatz}, J.~A., {Lo}, K.~Y., {et~al.} 2015,
  \href{http://adsabs.harvard.edu/abs/2015ApJ...800...26K}{\JournalTitle{\apj},
  800, 26}

\bibitem[{{Lauer} {et~al.}(2005){Lauer}, {Faber}, {Gebhardt}, {Richstone},
  {Tremaine}, {Ajhar}, {Aller}, {Bender}, {Dressler}, {Filippenko}, {Green},
  {Grillmair}, {Ho}, {Kormendy}, {Magorrian}, {Pinkney}, \&
  {Siopis}}]{laueretal2005}
{Lauer}, T.~R., {Faber}, S.~M., {Gebhardt}, K., {et~al.} 2005,
  \href{http://adsabs.harvard.edu/abs/2005AJ....129.2138L}{\JournalTitle{\aj},
  129, 2138}

\bibitem[{{Liu} {et~al.}(2009){Liu}, {Zakamska}, {Greene}, {Strauss}, {Krolik},
  \& {Heckman}}]{liuetal2009}
{Liu}, X., {Zakamska}, N.~L., {Greene}, J.~E., {et~al.} 2009,
  \href{http://adsabs.harvard.edu/abs/2009ApJ...702.1098L}{\JournalTitle{\apj},
  702, 1098}

\bibitem[{{Lo}(2005)}]{lo2005}
{Lo}, K.~Y. 2005,
  \href{http://adsabs.harvard.edu/abs/2005ARA%26A..43..625L}{\JournalTitle{\araa},
  43, 625}

\bibitem[{{Mandelbaum} {et~al.}(2009){Mandelbaum}, {Li}, {Kauffmann}, \&
  {White}}]{mandelbaumetal2009}
{Mandelbaum}, R., {Li}, C., {Kauffmann}, G., \& {White}, S.~D.~M. 2009,
  \href{http://adsabs.harvard.edu/abs/2009MNRAS.393..377M}{\JournalTitle{\mnras},
  393, 377}

\bibitem[{{Martini} {et~al.}(2013){Martini}, {Dicken}, \&
  {Storchi-Bergmann}}]{martinietal2013}
{Martini}, P., {Dicken}, D., \& {Storchi-Bergmann}, T. 2013,
  \href{http://adsabs.harvard.edu/abs/2013ApJ...766..121M}{\JournalTitle{\apj},
  766, 121}

\bibitem[{{Matthews} {et~al.}(1964){Matthews}, {Morgan}, \&
  {Schmidt}}]{matthewsetal1964}
{Matthews}, T.~A., {Morgan}, W.~W., \& {Schmidt}, M. 1964,
  \href{http://adsabs.harvard.edu/abs/1964ApJ...140...35M}{\JournalTitle{\apj},
  140, 35}

\bibitem[{{McConnell} \& {Ma}(2013)}]{mcconnellma2013}
{McConnell}, N.~J., \& {Ma}, C.-P. 2013,
  \href{http://adsabs.harvard.edu/abs/2013ApJ...764..184M}{\JournalTitle{\apj},
  764, 184}

\bibitem[{{Miyoshi} {et~al.}(1995){Miyoshi}, {Moran}, {Herrnstein},
  {Greenhill}, {Nakai}, {Diamond}, \& {Inoue}}]{miyoshietal1995}
{Miyoshi}, M., {Moran}, J., {Herrnstein}, J., {et~al.} 1995,
  \href{http://adsabs.harvard.edu/abs/1995Natur.373..127M}{\JournalTitle{\nat},
  373, 127}

\bibitem[{{Neufeld}(2000)}]{neufeld00a}
{Neufeld}, D.~A. 2000,
  \href{http://adsabs.harvard.edu/abs/2000ApJ...542L..99N}{\JournalTitle{\apjl},
  542, L99}

\bibitem[{{Neufeld} {et~al.}(1994){Neufeld}, {Maloney}, \&
  {Conger}}]{neufeldetal1994}
{Neufeld}, D.~A., {Maloney}, P.~R., \& {Conger}, S. 1994,
  \href{http://adsabs.harvard.edu/abs/1994ApJ...436L.127N}{\JournalTitle{\apjl},
  436, L127}

\bibitem[{{Onken} {et~al.}(2014){Onken}, {Valluri}, {Brown}, {McGregor},
  {Peterson}, {Bentz}, {Ferrarese}, {Pogge}, {Vestergaard}, {Storchi-Bergmann},
  \& {Riffel}}]{onkenetal14}
{Onken}, C.~A., {Valluri}, M., {Brown}, J.~S., {et~al.} 2014,
  \href{http://adsabs.harvard.edu/abs/2014ApJ...791...37O}{\JournalTitle{\apj},
  791, 37}

\bibitem[{{O'Sullivan} {et~al.}(2001){O'Sullivan}, {Forbes}, \&
  {Ponman}}]{osullivan01}
{O'Sullivan}, E., {Forbes}, D.~A., \& {Ponman}, T.~J. 2001,
  \href{http://adsabs.harvard.edu/abs/2001MNRAS.328..461O}{\JournalTitle{\mnras},
  328, 461}

\bibitem[{{Ott} {et~al.}(2013){Ott}, {Meier}, {McCoy}, {Peck}, {Impellizzeri},
  {Brunthaler}, {Walter}, {Edwards}, {Anderson}, {Henkel}, {Feain}, \&
  {Mao}}]{ott13}
{Ott}, J., {Meier}, D.~S., {McCoy}, M., {et~al.} 2013,
  \href{http://adsabs.harvard.edu/abs/2013ApJ...771L..41O}{\JournalTitle{\apjl},
  771, L41}

\bibitem[{{Paturel} {et~al.}(2003){Paturel}, {Petit}, {Prugniel}, {Theureau},
  {Rousseau}, {Brouty}, {Dubois}, \& {Cambr{\'e}sy}}]{patureletal2003}
{Paturel}, G., {Petit}, C., {Prugniel}, P., {et~al.} 2003,
  \href{http://adsabs.harvard.edu/abs/2003A%26A...412...45P}{\JournalTitle{\aap},
  412, 45}

\bibitem[{{Peng}(2007)}]{peng2007}
{Peng}, C.~Y. 2007,
  \href{http://adsabs.harvard.edu/abs/2007ApJ...671.1098P}{\JournalTitle{\apj},
  671, 1098}

\bibitem[{{Pesce} {et~al.}(2015){Pesce}, {Braatz}, {Condon}, {Gao}, {Henkel},
  {Litzinger}, {Lo}, \& {Reid}}]{pesce15}
{Pesce}, D.~W., {Braatz}, J.~A., {Condon}, J.~J., {et~al.} 2015,
  \href{http://adsabs.harvard.edu/abs/2015ApJ...810...65P}{\JournalTitle{\apj},
  810, 65}

\bibitem[{{Reid} {et~al.}(2009){Reid}, {Braatz}, {Condon}, {Greenhill},
  {Henkel}, \& {Lo}}]{reidetal2009}
{Reid}, M.~J., {Braatz}, J.~A., {Condon}, J.~J., {et~al.} 2009,
  \href{http://adsabs.harvard.edu/abs/2009ApJ...695..287R}{\JournalTitle{\apj},
  695, 287}

\bibitem[{{Reid} {et~al.}(2013){Reid}, {Braatz}, {Condon}, {Lo}, {Kuo},
  {Impellizzeri}, \& {Henkel}}]{reidetal2013}
---. 2013,
  \href{http://adsabs.harvard.edu/abs/2013ApJ...767..154R}{\JournalTitle{\apj},
  767, 154}

\bibitem[{{Reines} {et~al.}(2013){Reines}, {Greene}, \&
  {Geha}}]{reinesetal2013}
{Reines}, A.~E., {Greene}, J.~E., \& {Geha}, M. 2013,
  \href{http://adsabs.harvard.edu/abs/2013ApJ...775..116R}{\JournalTitle{\apj},
  775, 116}

\bibitem[{{Reines} \& {Volonteri}(2015)}]{reines15}
{Reines}, A.~E., \& {Volonteri}, M. 2015,
  \href{http://adsabs.harvard.edu/abs/2015ApJ...813...82R}{\JournalTitle{\apj},
  813, 82}

\bibitem[{{Robertson} {et~al.}(2006){Robertson}, {Hernquist}, {Cox}, {Di
  Matteo}, {Hopkins}, {Martini}, \& {Springel}}]{robertsonetal2006}
{Robertson}, B., {Hernquist}, L., {Cox}, T.~J., {et~al.} 2006,
  \href{http://adsabs.harvard.edu/abs/2006ApJ...641...90R}{\JournalTitle{\apj},
  641, 90}

\bibitem[{{Sadler} {et~al.}(1989){Sadler}, {Jenkins}, \&
  {Kotanyi}}]{sadleretal1989}
{Sadler}, E.~M., {Jenkins}, C.~R., \& {Kotanyi}, C.~G. 1989,
  \href{http://adsabs.harvard.edu/abs/1989MNRAS.240..591S}{\JournalTitle{\mnras},
  240, 591}

\bibitem[{{Sarzi} {et~al.}(2006){Sarzi}, {Falc{\'o}n-Barroso}, {Davies},
  {Bacon}, {Bureau}, {Cappellari}, {de Zeeuw}, {Emsellem}, {Fathi},
  {Krajnovi{\'c}}, {Kuntschner}, {McDermid}, \& {Peletier}}]{sarzietal2006}
{Sarzi}, M., {Falc{\'o}n-Barroso}, J., {Davies}, R.~L., {et~al.} 2006,
  \href{http://adsabs.harvard.edu/abs/2006MNRAS.366.1151S}{\JournalTitle{\mnras},
  366, 1151}

\bibitem[{{Shao} {et~al.}(2013){Shao}, {Kauffmann}, {Li}, {Wang}, \&
  {Heckman}}]{shaoetal13a}
{Shao}, L., {Kauffmann}, G., {Li}, C., {Wang}, J., \& {Heckman}, T.~M. 2013,
  \href{http://adsabs.harvard.edu/abs/2013MNRAS.436.3451S}{\JournalTitle{\mnras},
  436, 3451}

\bibitem[{{Shapiro} {et~al.}(2006){Shapiro}, {Cappellari}, {de Zeeuw},
  {McDermid}, {Gebhardt}, {van den Bosch}, \& {Statler}}]{shapiroetal06a}
{Shapiro}, K.~L., {Cappellari}, M., {de Zeeuw}, T., {et~al.} 2006,
  \href{http://adsabs.harvard.edu/abs/2006MNRAS.370..559S}{\JournalTitle{\mnras},
  370, 559}

\bibitem[{{Singh} {et~al.}(2013){Singh}, {van de Ven}, {Jahnke}, {Lyubenova},
  {Falc{\'o}n-Barroso}, {Alves}, {Cid Fernandes}, {Galbany},
  {Garc{\'{\i}}a-Benito}, {Husemann}, {Kennicutt}, {Marino}, {M{\'a}rquez},
  {Masegosa}, {Mast}, {Pasquali}, {S{\'a}nchez}, {Walcher}, {Wild}, {Wisotzki},
  \& {Ziegler}}]{singhetal13a}
{Singh}, R., {van de Ven}, G., {Jahnke}, K., {et~al.} 2013,
  \href{http://adsabs.harvard.edu/abs/2013A%26A...558A..43S}{\JournalTitle{\aap},
  558, A43}

\bibitem[{{Siopis} {et~al.}(2009){Siopis}, {Gebhardt}, {Lauer}, {Kormendy},
  {Pinkney}, {Richstone}, {Faber}, {Tremaine}, {Aller}, {Bender}, {Bower},
  {Dressler}, {Filippenko}, {Green}, {Ho}, \& {Magorrian}}]{siopisetal2009}
{Siopis}, C., {Gebhardt}, K., {Lauer}, T.~R., {et~al.} 2009,
  \href{http://adsabs.harvard.edu/abs/2009ApJ...693..946S}{\JournalTitle{\apj},
  693, 946}

\bibitem[{{Strauss} {et~al.}(2002){Strauss}, {Weinberg}, {Lupton}, {Narayanan},
  {Annis}, {Bernardi}, {Blanton}, {Burles}, {Connolly}, {Dalcanton}, {Doi},
  {Eisenstein}, {Frieman}, {Fukugita}, {Gunn}, {Ivezi{\'c}}, {Kent}, {Kim},
  {Knapp}, {Kron}, {Munn}, {Newberg}, {Nichol}, {Okamura}, {Quinn}, {Richmond},
  {Schlegel}, {Shimasaku}, {SubbaRao}, {Szalay}, {Vanden Berk}, {Vogeley},
  {Yanny}, {Yasuda}, {York}, \& {Zehavi}}]{straussetal2002}
{Strauss}, M.~A., {Weinberg}, D.~H., {Lupton}, R.~H., {et~al.} 2002,
  \href{http://adsabs.harvard.edu/abs/2002AJ....124.1810S}{\JournalTitle{\aj},
  124, 1810}

\bibitem[{{Sun} {et~al.}(2013){Sun}, {Greene}, {Impellizzeri}, {Kuo}, {Braatz},
  \& {Tuttle}}]{sunetal2013}
{Sun}, A.-L., {Greene}, J.~E., {Impellizzeri}, C.~M.~V., {et~al.} 2013,
  \href{http://adsabs.harvard.edu/abs/2013ApJ...778...47S}{\JournalTitle{\apj},
  778, 47}

\bibitem[{{Tarchi} {et~al.}(2007){Tarchi}, {Brunthaler}, {Henkel}, {Menten},
  {Braatz}, \& {Wei{\ss}}}]{tarchietal2007}
{Tarchi}, A., {Brunthaler}, A., {Henkel}, C., {et~al.} 2007,
  \href{http://adsabs.harvard.edu/abs/2007A%26A...475..497T}{\JournalTitle{\aap},
  475, 497}

\bibitem[{{Tarchi} {et~al.}(2003){Tarchi}, {Henkel}, {Chiaberge}, \&
  {Menten}}]{tarchietal2003}
{Tarchi}, A., {Henkel}, C., {Chiaberge}, M., \& {Menten}, K.~M. 2003,
  \href{http://adsabs.harvard.edu/abs/2003A%26A...407L..33T}{\JournalTitle{\aap},
  407, L33}

\bibitem[{{Toomre}(1964)}]{toomre64}
{Toomre}, A. 1964,
  \href{http://adsabs.harvard.edu/abs/1964ApJ...139.1217T}{\JournalTitle{\apj},
  139, 1217}

\bibitem[{{van den Bosch} \& {de Zeeuw}(2010)}]{vandenboschdezeeuw2010}
{van den Bosch}, R.~C.~E., \& {de Zeeuw}, P.~T. 2010,
  \href{http://adsabs.harvard.edu/abs/2010MNRAS.401.1770V}{\JournalTitle{\mnras},
  401, 1770}

\bibitem[{{van den Bosch} {et~al.}(2015){van den Bosch}, {Gebhardt},
  {G{\"u}ltekin}, {Y{\i}ld{\i}r{\i}m}, \& {Walsh}}]{vandenboschetal2015}
{van den Bosch}, R.~C.~E., {Gebhardt}, K., {G{\"u}ltekin}, K.,
  {Y{\i}ld{\i}r{\i}m}, A., \& {Walsh}, J.~L. 2015,
  \href{http://adsabs.harvard.edu/abs/2015ApJS..218...10V}{\JournalTitle{\apjs},
  218, 10}

\bibitem[{{van Dokkum} \& {Franx}(1995)}]{vandokkumfranx1995}
{van Dokkum}, P.~G., \& {Franx}, M. 1995,
  \href{http://adsabs.harvard.edu/abs/1995AJ....110.2027V}{\JournalTitle{\aj},
  110, 2027}

\bibitem[{{Vasudevan} \& {Fabian}(2007)}]{vasudevanfabian2007}
{Vasudevan}, R.~V., \& {Fabian}, A.~C. 2007,
  \href{http://adsabs.harvard.edu/abs/2007MNRAS.381.1235V}{\JournalTitle{\mnras},
  381, 1235}

\bibitem[{{Veilleux} \& {Osterbrock}(1987)}]{veilleuxosterbrock1987}
{Veilleux}, S., \& {Osterbrock}, D.~E. 1987,
  \href{http://adsabs.harvard.edu/abs/1987ApJS...63..295V}{\JournalTitle{\apjs},
  63, 295}

\bibitem[{{Walsh} {et~al.}(2013){Walsh}, {Barth}, {Ho}, \&
  {Sarzi}}]{walshetal2013}
{Walsh}, J.~L., {Barth}, A.~J., {Ho}, L.~C., \& {Sarzi}, M. 2013,
  \href{http://adsabs.harvard.edu/abs/2013ApJ...770...86W}{\JournalTitle{\apj},
  770, 86}

\bibitem[{{Walsh} {et~al.}(2012){Walsh}, {van den Bosch}, {Barth}, \&
  {Sarzi}}]{walshetal12a}
{Walsh}, J.~L., {van den Bosch}, R.~C.~E., {Barth}, A.~J., \& {Sarzi}, M. 2012,
  \href{http://adsabs.harvard.edu/abs/2012ApJ...753...79W}{\JournalTitle{\apj},
  753, 79}

\bibitem[{{Wardle} \& {Yusef-Zadeh}(2012)}]{wardle12}
{Wardle}, M., \& {Yusef-Zadeh}, F. 2012,
  \href{http://adsabs.harvard.edu/abs/2012ApJ...750L..38W}{\JournalTitle{\apjl},
  750, L38}

\bibitem[{{Wiklind} \& {Henkel}(2001)}]{wiklind01}
{Wiklind}, T., \& {Henkel}, C. 2001,
  \href{http://adsabs.harvard.edu/abs/2001A%26A...375..797W}{\JournalTitle{\aap},
  375, 797}

\bibitem[{{Wrobel} \& {Heeschen}(1991)}]{wrobelheeschen1991}
{Wrobel}, J.~M., \& {Heeschen}, D.~S. 1991,
  \href{http://adsabs.harvard.edu/abs/1991AJ....101..148W}{\JournalTitle{\aj},
  101, 148}

\bibitem[{{Yee}(1980)}]{yee1980}
{Yee}, H.~K.~C. 1980,
  \href{http://adsabs.harvard.edu/abs/1980ApJ...241..894Y}{\JournalTitle{\apj},
  241, 894}

\bibitem[{{York} {et~al.}(2000)}]{yorketal2000}
{York}, D.~G., {et~al.} 2000,
  \href{http://adsabs.harvard.edu/abs/2000AJ....120.1579Y}{\JournalTitle{\aj},
  120, 1579}

\bibitem[{{Zakamska} {et~al.}(2003)}]{zakamskaetal2003}
{Zakamska}, N.~L., {et~al.} 2003,
  \href{http://adsabs.harvard.edu/abs/2003AJ....126.2125Z}{\JournalTitle{\aj},
  126, 2125}

\bibitem[{{Zhang} {et~al.}(2010){Zhang}, {Henkel}, {Guo}, {Wang}, \&
  {Fan}}]{zhangetal2010}
{Zhang}, J.~S., {Henkel}, C., {Guo}, Q., {Wang}, H.~G., \& {Fan}, J.~H. 2010,
  \href{http://adsabs.harvard.edu/abs/2010ApJ...708.1528Z}{\JournalTitle{\apj},
  708, 1528}

\bibitem[{{Zhang} {et~al.}(2012){Zhang}, {Henkel}, {Guo}, \& {Wang}}]{zhang12}
{Zhang}, J.~S., {Henkel}, C., {Guo}, Q., \& {Wang}, J. 2012,
  \href{http://adsabs.harvard.edu/abs/2012A%26A...538A.152Z}{\JournalTitle{\aap},
  538, A152}

\bibitem[{{Zhu} {et~al.}(2011){Zhu}, {Zaw}, {Blanton}, \&
  {Greenhill}}]{zhuetal2011}
{Zhu}, G., {Zaw}, I., {Blanton}, M.~R., \& {Greenhill}, L.~J. 2011,
  \href{http://adsabs.harvard.edu/abs/2011ApJ...742...73Z}{\JournalTitle{\apj},
  742, 73}

\end{thebibliography}
\end{document}